\documentclass[11pt,a4paper]{article}
\pdfoutput=1
\usepackage{jheppub}

\usepackage{float}
\usepackage{amsmath}
\usepackage{cleveref}
\usepackage{xcolor}
\usepackage{caption}
\usepackage{subcaption}
\usepackage{physics}

\crefname{equation}{eqn.}{eqns.}
\Crefname{equation}{Eqn.}{Eqns.}
\crefname{figure}{fig.}{figs.}
\Crefname{figure}{Fig.}{Figs.}
\newcommand{\I}{\mathcal{I}}
\newcommand{\td}{\text{d}}

\usepackage{mathdots}
\usepackage{arydshln}

\title{Optimisation of Thimble Simulations and Quantum Dynamics of Multiple Fields in Real Time}

\author[a]{Simon Woodward,}
\author[a]{Paul M. Saffin,}
\author[b]{Zong-Gang Mou,}
\author[b]{Anders Tranberg}

\affiliation[a]{School of Physics and Astronomy, University Park, University of Nottingham,\\ Nottingham NG7 2RD, United Kingdom}
\affiliation[b]{Faculty of Science and Technology, University of Stavanger, 4036 Stavanger, Norway}

\emailAdd{simon.woodward1@nottingham.ac.uk}
\emailAdd{paul.saffin@nottingham.ac.uk} 
\emailAdd{zonggang.mou@uis.no}
\emailAdd{anders.tranberg@uis.no}

\abstract{We apply the Generalised Thimble approach to the computation of exact path integrals and correlators in real-time quantum field theory. We first investigate the details of the numerical implementation and ways of optimizing the algorithm. We subsequently apply the method to an interacting two-field system in 0+1 dimensions, illustrating the scope for addressing realistic physical processes using real-time Generalised Thimble computations.}

\begin{document}

\maketitle

%%%%%%%%%%%%%%%%%%%%%%%%%%%%%%%%%%%%%%%%%%%%%%%%%%%%%

\section{Introduction}
\label{sec:Intro}
The time evolution of quantum systems is represented by non-equal time correlators of the form
\begin{equation}
\label{eqn:expectation}
    \langle \hat{\mathcal{O}}(t_1) \hat{\mathcal{O}}(t_2) \rangle = \text{Tr}\left( \hat{\mathcal{O}}(t_1) \hat{\mathcal{O}}(t_2) \hat{\rho}\right),
\end{equation}
where $\hat{\rho}$ represents the density matrix. Separating out the time-evolution operators $U(t)=e^{i\hat{H}t}$, allows us to rewrite these quantum expectation values as path integrals in real time, with the field variable living on the Keldysh contour. This contour runs from the initial condition at $t=0$ to some finite time larger than both $t_1$ and $t_2$, and returns to $t=0$. In this Heisenberg picture, the density matrix simply represents the initial state at $t=0$. 
In equilibrium systems, this density matrix may also be written in terms of the Hamiltonian $\hat{H}$ as $\hat{\rho} = e^{-\beta \hat{H}}/\text{Tr}(e^{-\beta \hat{H}})$. Further identifying the inverse temperature $\beta = 1/T$ with an imaginary time, this density matrix is formally equivalent to time evolution along the Keldysh contour extended to $-i\beta$, and may be included in the path integral formulation that way.

Fully non-perturbative evaluation of equal-time ($t_1=t_2$) correlators in thermal equilibrium is by now routine through the lattice discretization of field theory systems, and the application of numerical importance (Monte-Carlo, MC) sampling. This works, because the weights of paths in the path integral, $e^{iS}$, are real (and positive) when evaluated only on the imaginary part of the time contour, $e^{iS}\rightarrow e^{-S_E}$ and these may therefore be sampled as a probability distribution. The statistical averaging converges well, since paths with large Euclidean action are exponentially suppressed. 

Unfortunately, when evaluating non-equal time or non-equilibrium correlators, the time contour is no longer purely imaginary, the weights are complex and oscillating rather than real, and standard importance sampling techniques no longer apply. This is known as the "sign problem" of real-time lattice field theory.  

Over the past several years, possible avenues to solving this problem have been explored. Some of these involve allowing the real field variables, for the purpose of evaluating the path integral only, to take on complex values (see \cite{Berges:2006xc} for an early work, and \cite{Alexandru:2020wrj} for a brief review). This renders the action complex and can in some cases make the path integral better convergent. Recently, the use of Picard-Lefshetz Thimbles or Generalised Thimble methods have successfully mitigated the sign problem for fully real-time processes, but still at a considerable computational cost \cite{mou2019real, mou2019quantum, tanizaki2014real}. 

The central insight is that since the action is a smooth (typically polynomial) function of the field variables, the integral over these field variables is unchanged by deforming the integration region from the real axis to some other contour in the complex plane. An efficient evaluation becomes a question of finding the optimal (or just a good) deformation of $\mathbb{R}^n$ into $\mathbb{C}^n$, where $n$ is the number of field variables to be integrated over\footnote{We will work on a finite space-time lattice, and so $n$ is finite, although often large.}. Picard-Lefshetz thimbles or Generalised Thimble methods provide an algorithm for doing this. 

Much attention has been given to finite density problems in QCD, where the sign problem may likely be alleviated through complexification of the variables using Complex Langevin dynamics (see for instance \cite{Aarts:2008rr,Sexty:2013ica}) and recenlty through the method of thimbles \cite{Cristoforetti:2012su,Cristoforetti:2013wha}. 
The sign problem is even more severe for the real-time evolution out of equilibrium, but we were able to demonstrate that also for this case, the method of thimbles provides a significant improvement \cite{mou2019real}. In short, one may apply standard importance sampling to the initial density matrix for the field variables at $t=0$, and subsequently evaluate integrals over the $t>0$ variables using the thimble methods. This works for initial states corresponding to positive definite Wigner functions.

As the number of field variables increases, the computational effort of evaluating correlators grows exponentially due to the sign problem. In principle, complexifying the variables by means of the Picard-Lefshetz Thimbles and Generalised Thimble methods resolves this problem, but the computational cost is still substantial and grows as a power of the number of field variables. Hence, full field theory in 3+1 dimensions for any useful physical time-scale requires further analysis, diagnostics and optimisation of the bottlenecks of the numerical implementation. In the first part of the present paper, we will perform such an analysis in the context of field theory in 0+1 dimensions. This is a numerically manageable system, allowing us to better investigate the space of physical parameters as well as parameters of the numerical implementation. It also has the advantage that direct computation of the evolution using standard quantum mechanical evolution is possible to do very cheaply, providing something to compare our results to. In the second part, we will implement the Generalised Thimble method to a system of two interacting scalar fields. Mixing and decay of interacting fields in real time is an obvious (current and future) application of the Generalised Thimbles method. In addition to confirming the applicability and accuracy of the method, we will be able to assess what size lattice, what amount of MC sampling, and consequently how much computing time is required to convincingly tackle relevant physical processes. 

The paper is structured as follows: In Section 2, we present the Lefshetz Thimble and Generalised Thimble methods in some detail and offer a toy model example. We also set up the types of initial conditions, we will be considering. In Section 3 we dive deeper into the technology of the numerical implementation and identify the primary bottlenecks. We then investigate ways of tuning the numerics for optimal convergence. In Section 4, we introduce a second field, and present results for the correlator for the decoupled case, for the case when the two fields mix, and for when they are quartically coupled allowing one species to decay into the other. We conclude in section 5.

Since we are in effect working in quantum mechanics (field theory in 0+1 dimensions), we in Appendix A present the (numerically much simpler) standard quantum mechanical method we will use for comparison. 

%%%%%%%%%%%%%%%%%%%%%%%%%%%%%%%%%%%%%%%%%%%%%%%%%%%%%

\section{Path Integrals in the complex plane}
\label{sec:thimble}

Consider a path integral of the form 
\begin{equation}
    \label{eqn:path_integral}
    A = \int_{\mathbb{R}^n} \mathcal{D}\varphi \; e^{-\I(\varphi)},
\end{equation}
with real, scalar variables $\varphi$. It is implied that there is a finite number $n$ of variables, so that $\mathcal{D}\varphi=\Pi_n d\varphi$. The action $-\I=iS$ is a function of all the $\varphi$, and most often imaginary, meaning the integrand is oscillatory with a constant unit amplitude but variable phase\footnote{Note that in the following, we are considering the object $\I=-iS$ rather than $S$ itself.}. This variable phase is the root cause of the "sign problem", and makes the integral difficult to evaluate even using numerical methods. A solution to this problem is provided by a multidimensional version of Cauchy's theorem. By promoting the real variables $\varphi$ to complex variables, denoted $\phi$, the integration manifold $\mathbb{R}^n$ can be deformed into an $n$-dimensional manifold in $\mathbb{C}^n$ without altering the value of the integral. The task is to select a manifold, where the integrand has no, or at least better behaved, oscillations. We note that although this amounts to deforming the integration regions $\mathbb{R}$ for each of the $\varphi$ into the complex plane, the optimal common $n$-dimensional manifold may not necessarily follow from deforming the domain of each $\varphi$ independently. 

%%%%%%%%%%%%%%%%%

\subsection{Lefschetz Thimbles}
\label{Lefschetz}

Picard-Lefschetz theory provides a flow equation to find an appropriate manifold, known as a Lefshetz Thimble. 
Given that the action is $\mathcal{I}(\{\phi_j\})$, given some initial values for all the variable $\phi_j$, we can "flow" them in (non-physical) time $\tau$, according to
\begin{equation}
    \label{eqn:flow}
    \frac{\text{d}\phi_j}{\text{d}\tau} = \overline{\frac{\partial \I}{\partial \phi_j}}.
\end{equation}
This is a coupled set of equations for all the variables $\phi_j$, for which the right-hand side is in general not real. So given an initial set of real values for all the $\varphi_j$ (a real configuration), we can flow them in $\tau$ to a corresponding set of complex values $\phi_j$ (a complex configuration). And we may evaluate the integrand $e^{-\mathcal{I}}$ at that configuration rather than the original real-valued one. 

One property of this procedure is that the classical solution (or in this context, the critical point in the space of field configurations) satisfying
\begin{equation}
    \label{eqn:critical}
    \left. \frac{\partial \I}{\partial \phi_j} \right|_{critical} = 0 ,
\end{equation}
is a fixed point of the flow, and the corresponding integrand contribution is unchanged. For all other configurations, we see that as the flow proceeds, the action changes according to
\begin{equation}
\label{eqn:damp}
\begin{split}
    \frac{\partial \I}{\partial \tau} &= \sum_j \frac{\partial \I}{\partial \phi_j}\frac{\partial \phi_j}{\partial \tau} 
    = \sum_j \frac{\partial \I}{\partial \phi_j} \overline{\frac{\partial S}{\partial \phi_j}} 
    = \sum_j \left| \frac{\partial \I}{\partial \phi_j} \right|^2.
\end{split}
\end{equation}
This quantity is explicitly real, and consequently, the imaginary part of $\mathcal{I}$ is unchanged during the flow while a positive real part is acquired, exponentially suppressing the oscillating integrand contribution of the configuration. In principle, flowing to $\tau \rightarrow \infty$, the oscillations are removed completely, but this is rarely possible.

This procedure ensures that the full path integral over the entire $\mathbb{R}^n$ with equal amplitude integrands everywhere, now reduces to integrals over localised regions (thimbles) in configuration space close to the classical configurations (critical points). Care must be taken to include the contributions from all such thimbles, one for each classical solution (or stationary point) of the action. 

\begin{figure}[htb!]
    \centering
    \begin{subfigure}[b]{0.42\textwidth}
        \centering
        \includegraphics[width = \textwidth]{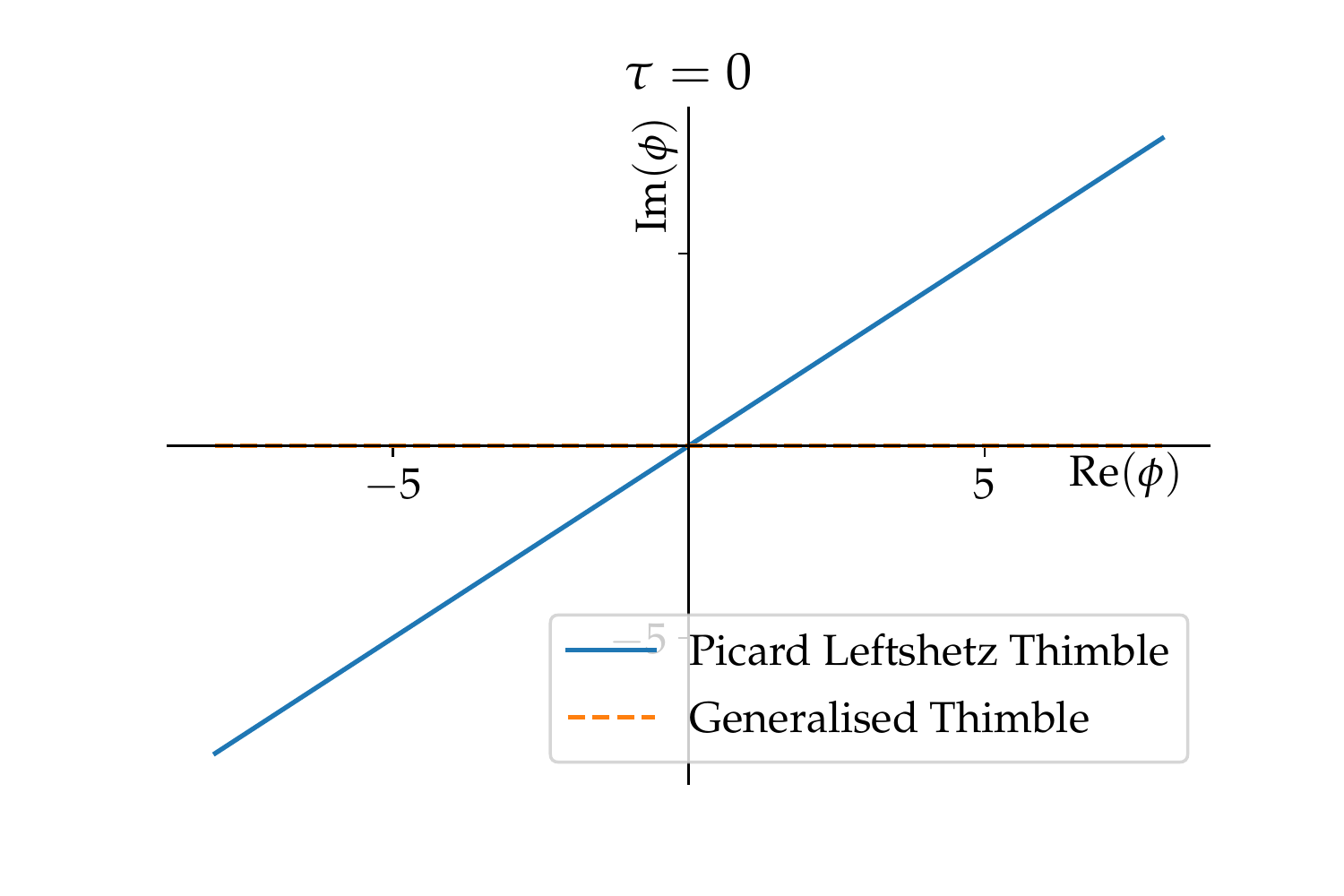}
        \caption{}
    \end{subfigure}
        \begin{subfigure}[b]{0.42\textwidth}
        \centering
        \includegraphics[width = \textwidth]{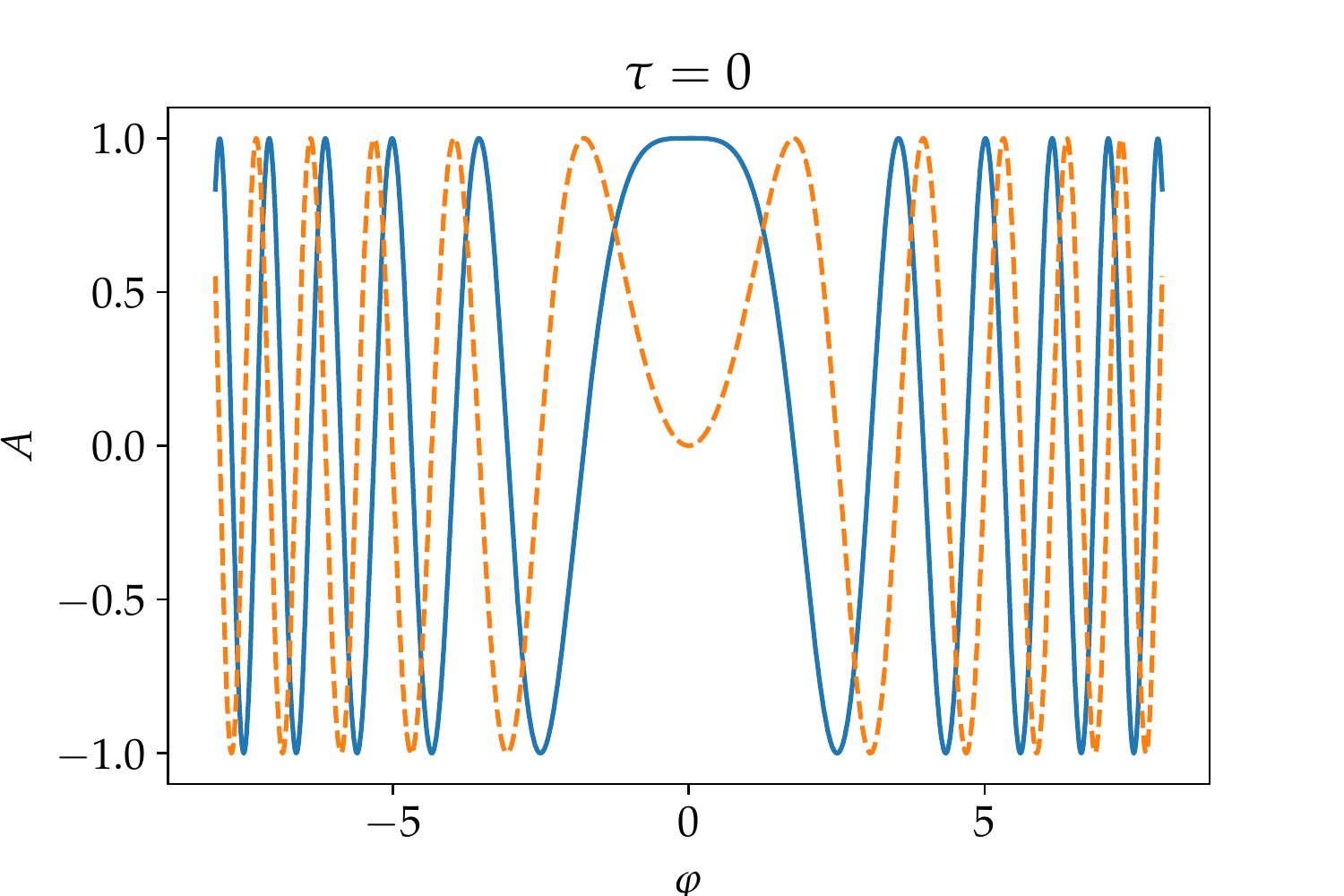}
        \caption{}
    \end{subfigure}
        \begin{subfigure}[b]{0.42\textwidth}
        \centering
        \includegraphics[width = \textwidth]{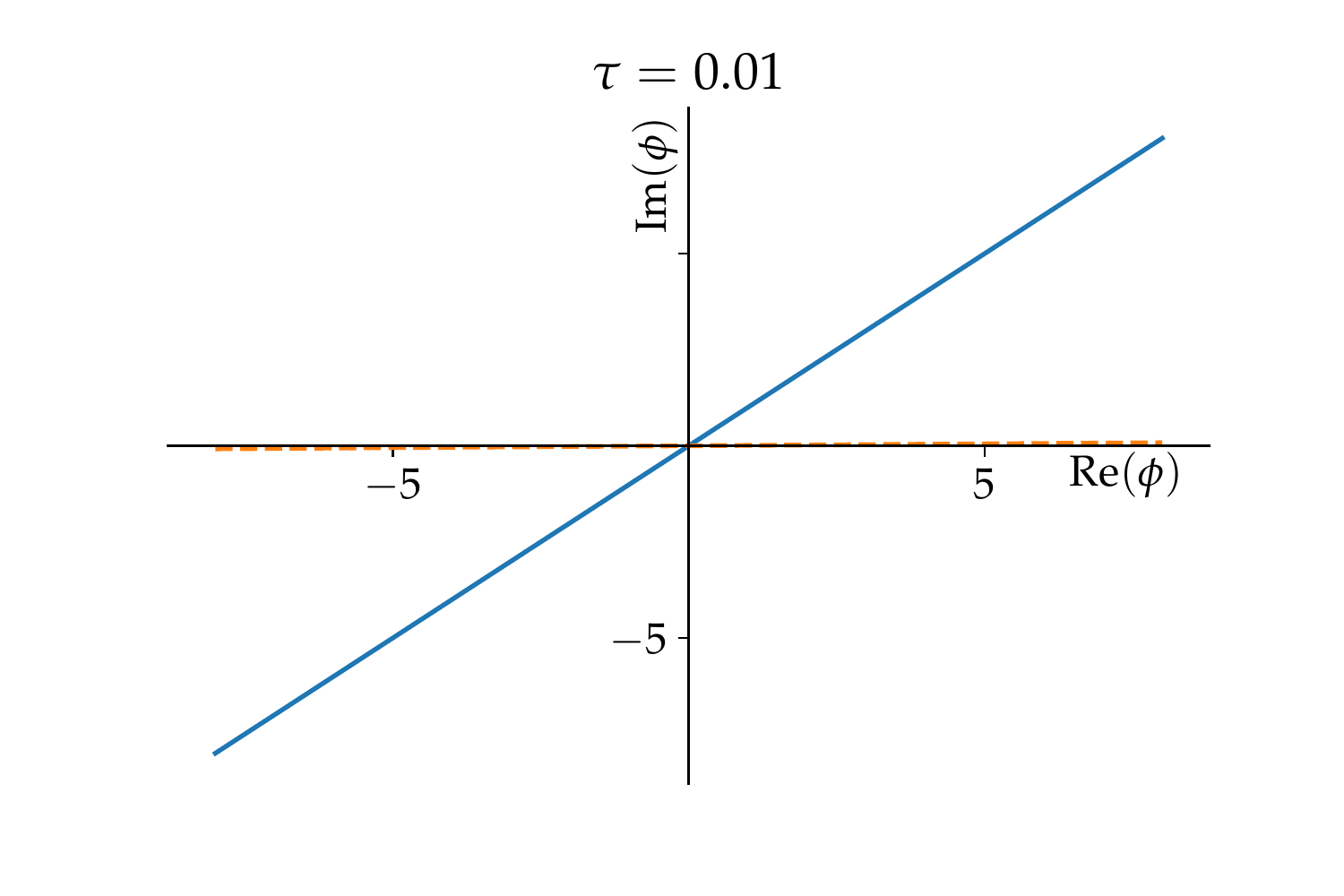}
        \caption{}
    \end{subfigure}
        \begin{subfigure}[b]{0.42\textwidth}
        \centering
        \includegraphics[width = \textwidth]{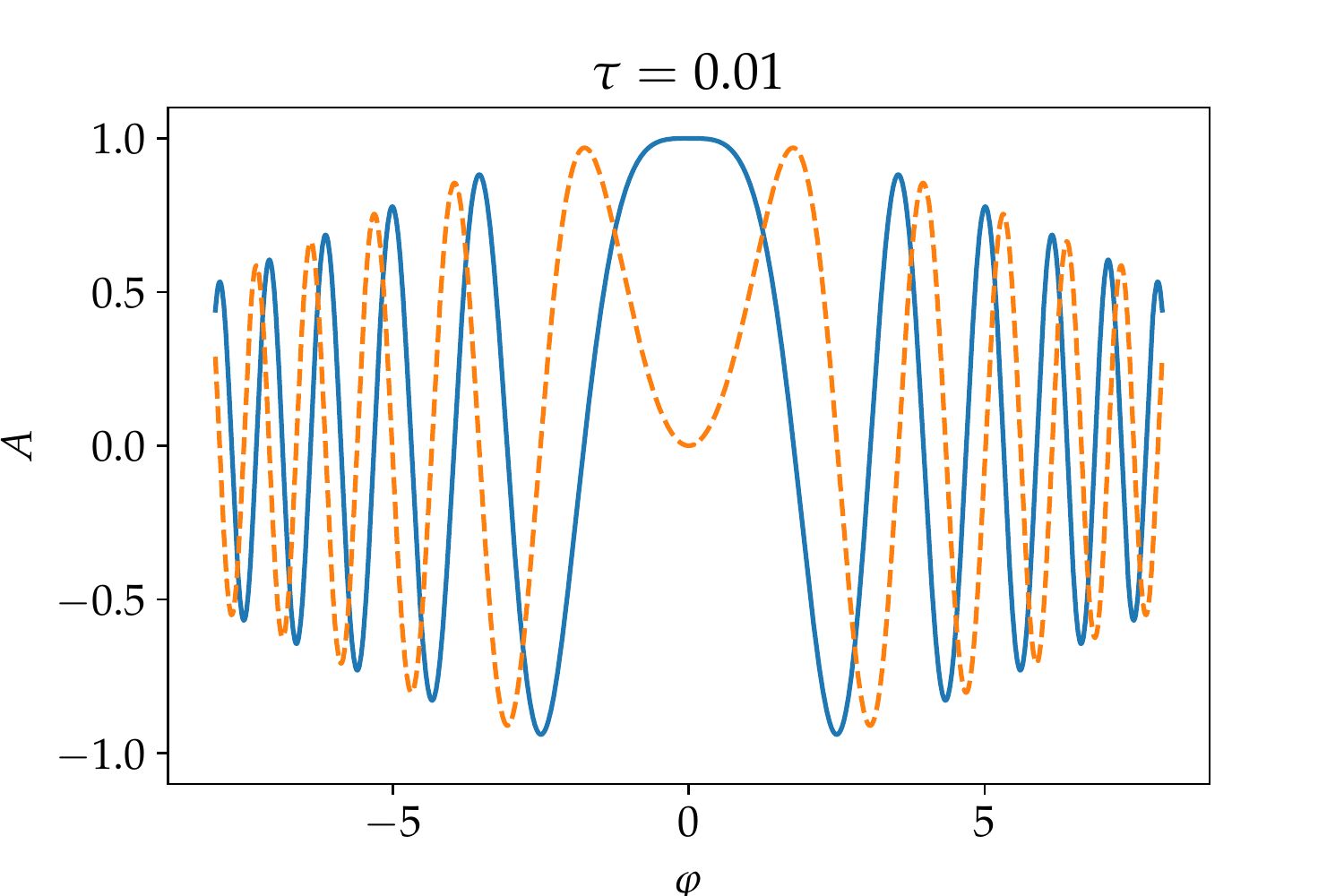}
        \caption{}
    \end{subfigure}
        \begin{subfigure}[b]{0.42\textwidth}
        \centering
        \includegraphics[width = \textwidth]{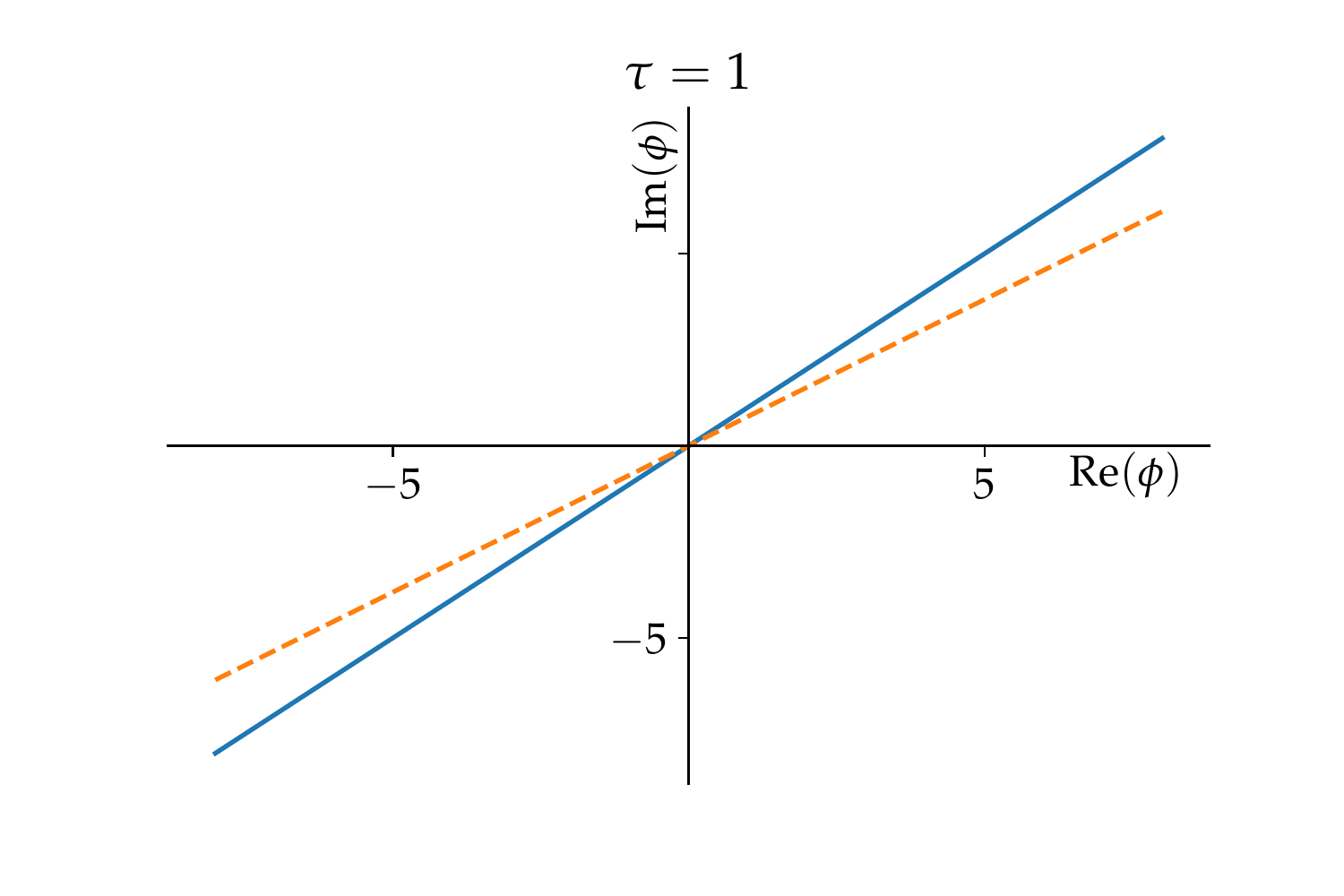}
        \caption{}
    \end{subfigure}
        \begin{subfigure}[b]{0.42\textwidth}
        \centering
        \includegraphics[width = \textwidth]{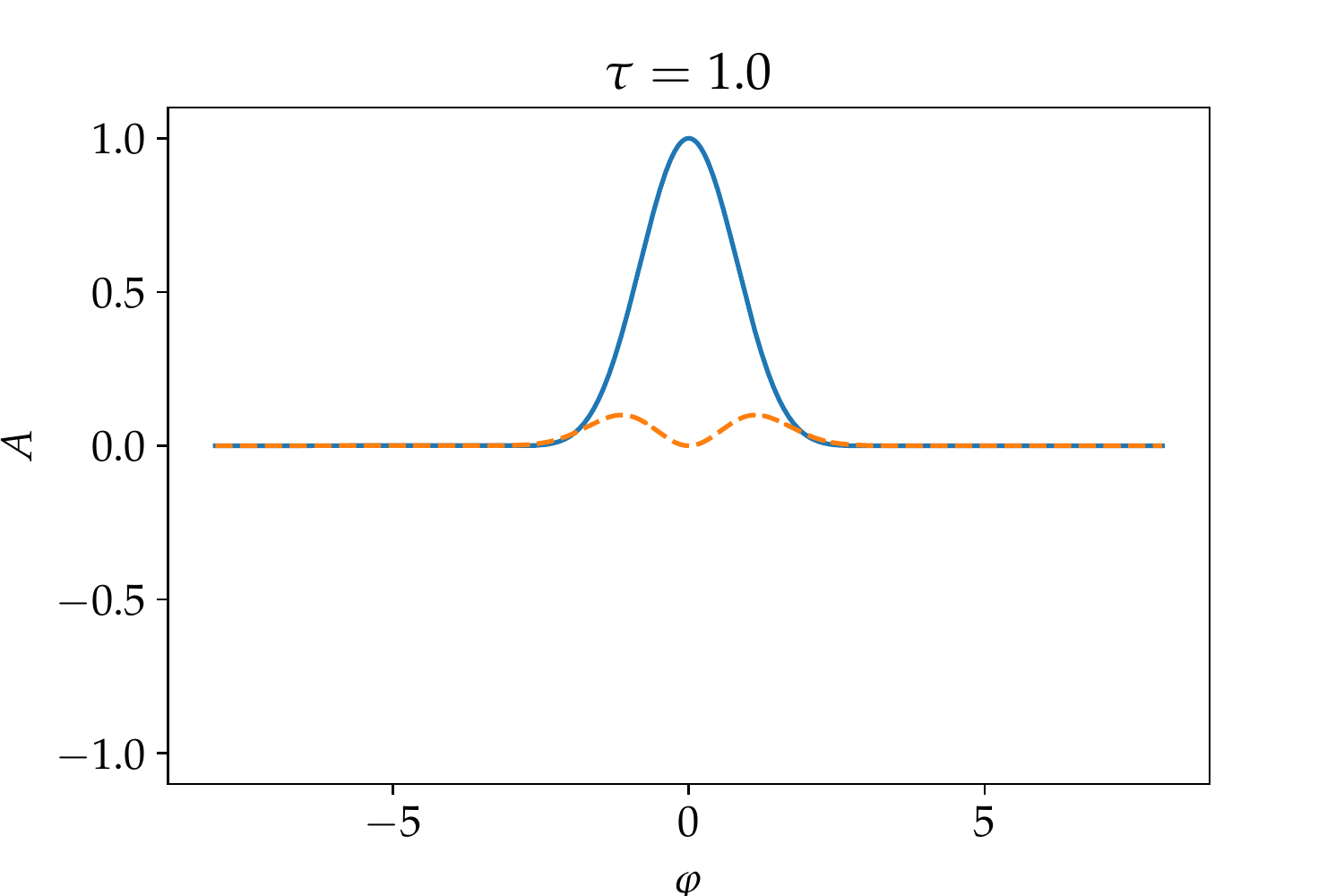}
        \caption{}
    \end{subfigure}
\begin{subfigure}[b]{0.42\textwidth}
        \centering
        \includegraphics[width = \textwidth]{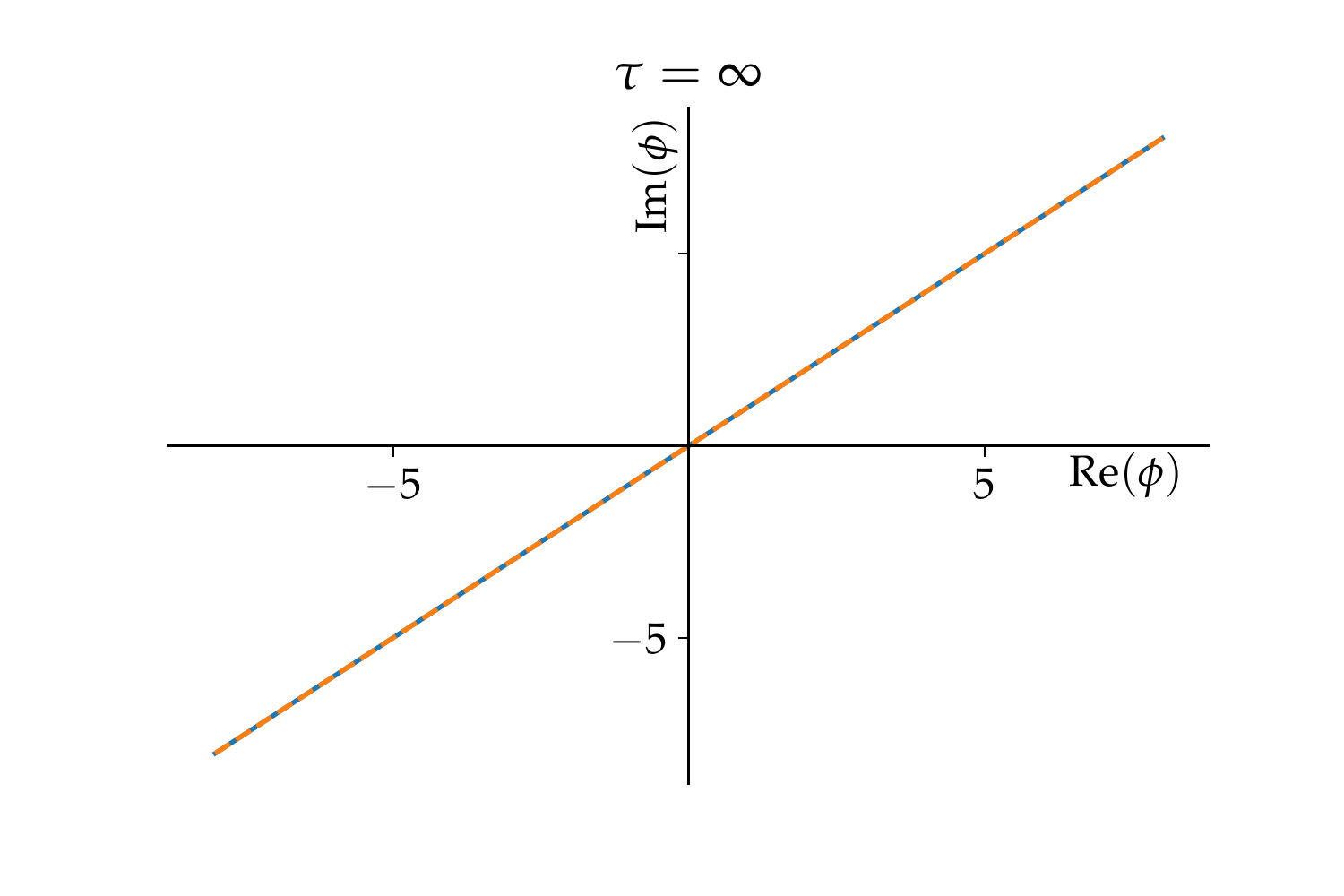}
        \caption{}
    \end{subfigure}
        \begin{subfigure}[b]{0.42\textwidth}
        \centering
        \includegraphics[width = \textwidth]{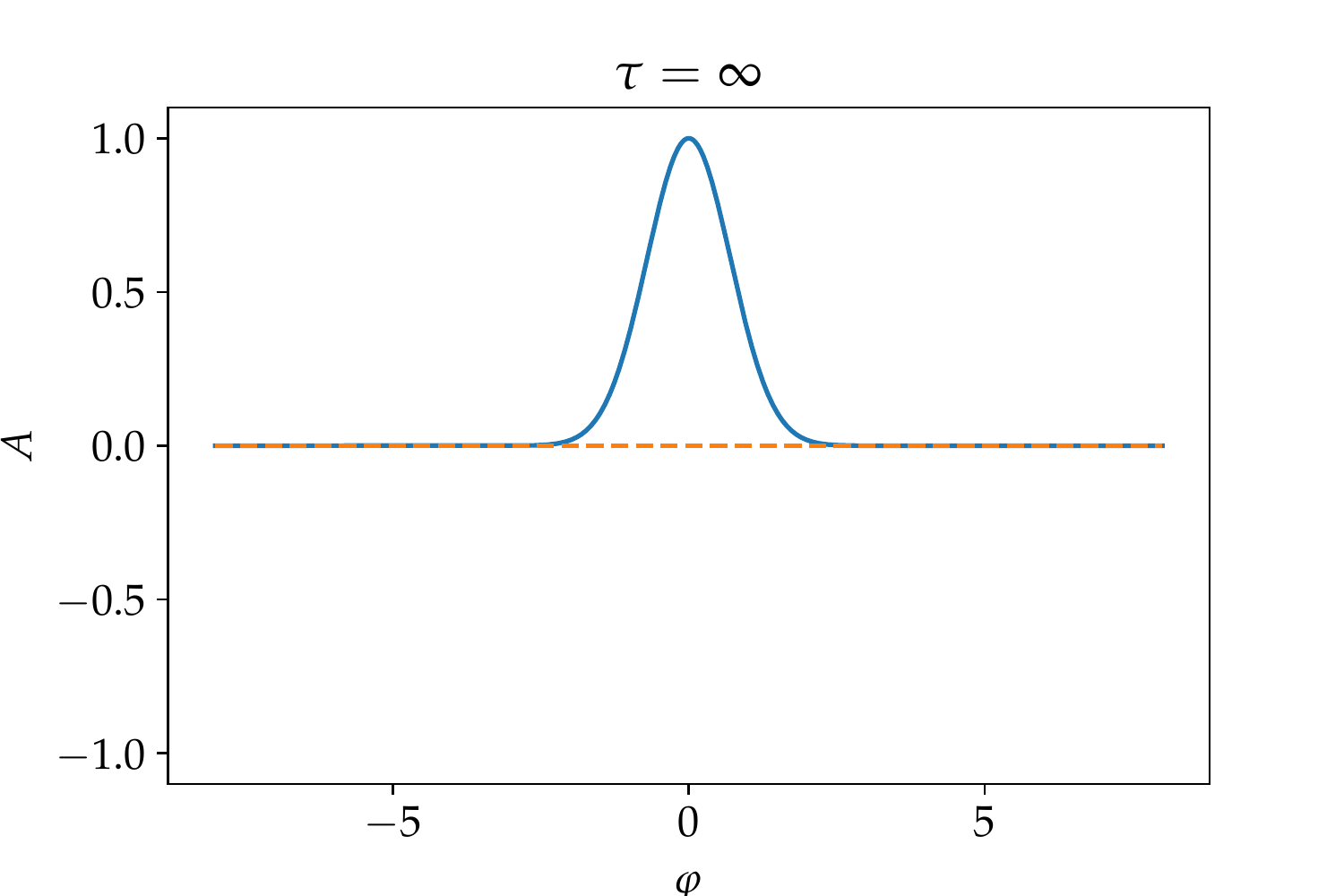}
        \caption{}
    \end{subfigure}
    \caption{The thimbles (left) and the real and imaginary parts of the integrand along the generalized thimble (right), $A = \exp[-\I(\phi)]$, where $\I(\phi) = -\frac{1}{2}i\phi^2$, for a number of different flow times $\tau$.}
    \label{fig:airy_demo}
\end{figure}

To illustrate the procedure for a very simple example, consider $\I = - \frac{1}{2}i \varphi^2$, with a single variable, $n=1$. The flow equation (\ref{eqn:flow}) gives 
\begin{equation}
    \frac{\td \phi}{\td \tau} = \overline{-i\phi}.
\end{equation}
Writing $\phi = a + ib$
\begin{equation}
    \dot{a} + i\dot{b} = i(a - ib),
\end{equation}
and splitting into real and imaginary parts gives
\begin{equation}
    \begin{split}
        \dot{a} = b, \quad   \dot{b} = a \rightarrow  \ddot{a} = a, \quad \ddot{b} = b.
    \end{split}
\end{equation}
The critical point is $\phi=0$, and for any real initial value $a_0$, we find
\begin{equation}
\label{eqn:a_b}
\begin{split}
    a = a_0\cosh(\tau), \qquad b = a_0\sinh(\tau).
\end{split}
\end{equation}
i.e. a straight line in the $\phi$-plane, of gradient $\cosh{\tau}$.
The thimble follows from taking $\tau\rightarrow \infty$, where $a=b$, as illustrated in Figure \ref{fig:airy_demo}. In that figure we also show the integrand $A=e^{-\mathcal{I}}$, which in the limit is just $e^{-a^2}$, rather than the original ($\tau=0$) integrand value $e^{ia_0^2/2}$. 

%%%%%%%%%%%%

\subsection{Generalised Thimble Method}
\label{sec:GenThimbles}

%%%%%%%%%%%

For multi-variable systems it is seldom possible to analytically find the Lefshetz thimble. Instead, a numerical solution of the flow equation is required, which in practice means setting a finite maximum flow time $\tau_{max}$. In this way, one may flow the initial manifold $\mathcal{M}_0 = \mathbb{R}^n$ (real variables $\varphi_j$) to some other, complex manifold $\mathcal{M}_\tau$ (variables $\phi_j$). The transformation between the two is encoded in the Jacobian
\begin{equation}
    \label{eqn:jacobian_definition}
    J_{ij} = \frac{\partial \phi_i}{\partial \varphi_j} ,
\end{equation}
so that the path integral may be written
\begin{equation}
    \label{eqn:jacobian_integral}
    \begin{split}
        \int_{\mathcal{M}_0} \mathcal{D}\varphi \; e^{-\I(\varphi)} &= \int_{\mathcal{M}_\tau} \mathcal{D}\phi \; e^{-\I(\phi)} = \int_{\mathcal{M}_0} \mathcal{D}\varphi \; \det(J) e^{-\I(\phi)}.
    \end{split}
\end{equation}
This Jacobian has its own flow equation
\begin{equation}
    \label{eqn:Jacobian_calculation}
    \frac{\td}{\td \tau} J_{ij} = \sum_s \overline{\frac{\partial^2 \I}{\partial \phi_i \partial \phi_s} J_{sj}},
\end{equation}
with $J = \mathbb{I}$ at $\tau=0$. In Figure \ref{fig:airy_demo}, left-hand panel, we show the Generalised Thimble (orange) and the asymptotic thimble (blue), while the right-hand panel shows the path integral integrand $A$ (real/blue and imaginary/orange parts) at different flow times. For $\tau=0$, the field only takes real values, and $A$ is strongly oscillating. Flowing to $\tau=0.01$ makes little difference, as $\phi$ still takes only values close to the real axis, and the $A$ is somewhat damped, but still very much an oscillating function. The situation is very different for $\tau=1$, where $A$ is strongly suppressed away from $\phi=0$, and $\phi$ takes values close to, but not yet on the thimble. As $\tau\rightarrow \infty$, the integrand $A$ is purely real and gaussian near the critical point $\phi=0$. 

%%%%%%%%%%%%%%%%%%%%%%%%%%%%%%%%%%%%%%%%%%%%%%%%%%%%%

\subsection{The field theory path integral}
\label{sec:Path_integral}

%%%%%%%%%%%%%%%%%%%%%%%%%%%%%%%%%%%%%%%%%%%%%%%%%%%%

\begin{figure}
    \centering
    \includegraphics[scale = 1.2]{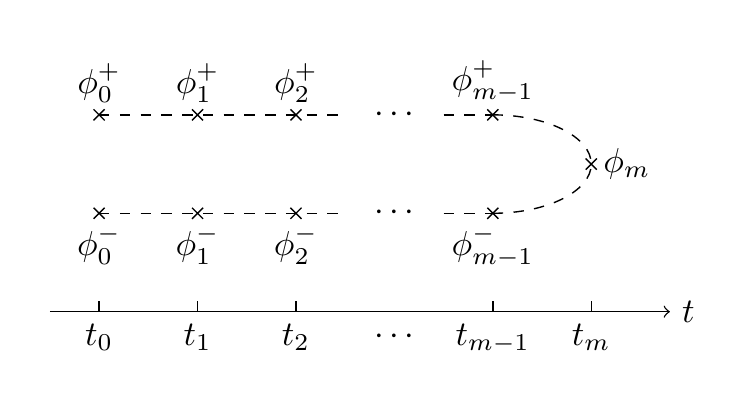}
    \caption{The discretized Schwinger-Keldysh contour.}
    \label{fig:contour}
\end{figure}
The path integral formalism on the Schwinger-Keldysh contour can be used to evaluate expressions of the form
\begin{equation}
\label{eqn:path_integral_expectation}
    \left\langle \hat{\mathcal{O}}(t) \right \rangle = \text{Tr}\left(\hat{\mathcal{O}}(t) \hat{\rho}(t_0) \right) = C \int \mathcal{D} \phi \; \mathcal{O}(t) \left \langle \phi^+_0; t_0| \hat{\rho} | \phi_0^-; t_0 \right\rangle \exp{\left(\frac{i}{\hbar}\int \text{d}t L\right)}.
\end{equation}
The top (bottom) branch field variables are indicated with $+$ ($-$), $\hat{\rho}$ is the density matrix, of which the matrix elements are taken at the initial time $t_0$. The Lagrangian $L$ is a function of $\phi^+_n$, $\phi^-_n$, and we will in the following have in mind, that time is discretized as displayed in Figure \ref{fig:contour}, with a finite number of time steps $N_{tot} = 2m + 1$. $C$ is a constant.

The initial state is left unspecified, and our contour does not have an imaginary time extension. Both the top and bottom branches are exactly on the real axis and are only separated here for clarity. 

Keeping in mind our discussion above and Eq. (\ref{eqn:jacobian_integral}), we may rewrite this expression as 
\begin{equation}
\label{eqn:observable_samples}
    \left \langle \hat{\mathcal{O}}(t) \right \rangle = \frac{\int \mathcal{D} \phi \; e^{-\mathcal{I}(\phi)} \hat{\mathcal{O}}}{\int \mathcal{D} \phi \; e^{-\mathcal{I}(\phi)}} = \frac{\left \langle e^{-i \text{Im}[\mathcal{I}(\phi)] + i\arg[\det(J)]} \hat{\mathcal{O}} \right \rangle_P}{\left \langle e^{-i \text{Im}[\mathcal{I}(\phi)] + i\arg[\det(J)]}\right \rangle_P},
\end{equation}
where the expectation values are evaluated over a distribution $P$, defined as
\begin{equation}
    P(\phi) = e^{-\text{Re}[\mathcal{I}(\phi)] + \ln|\det(J)|}.
\end{equation}

%%%%%%%%%%%%%%%%%%%%%%%%%%%%%

\subsection{Initial density matrix for $n$-particle states}
\label{sec:n_particle_density_matrix}

%%%%%%%%%%%%%%%%%%%%%%%%%%%%%

\begin{figure}
    \centering
    \includegraphics[scale = 1.2]{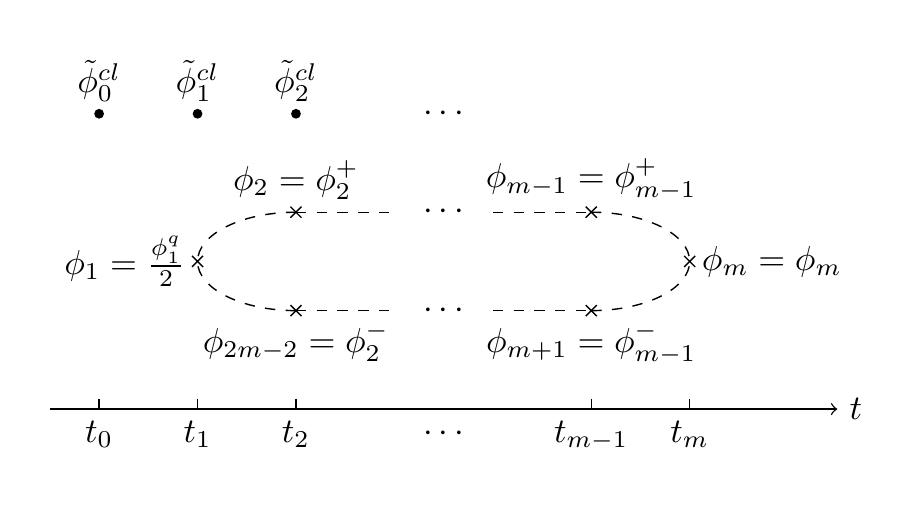}
    \caption{Reduced contour by integrating out the initial conditions. $\Tilde{\phi}_n^{cl}$ represents the solution to the equation of motion for $\phi^{cl}$ for a given set of initial conditions.}
    \label{fig:reduced_contour}
\end{figure}

The initial conditions are defined at the initial time $t=0$ through the variables $\phi_0$, $\dot{\phi}_0$, which in discretized time involves $\phi_0$ and $\phi_1$, $\dot{\phi}_0=(\phi_1-\phi_0)/dt$. The appropriate set of field variables to sample are the Keldysh basis \cite{aarts1998classical, fukuma2017parallel}, 
\begin{eqnarray}
    \phi^{cl}_n &= \frac{1}{2}(\phi^+_n + \phi^-_n), \qquad
    \phi^q_n &= \phi^+_n - \phi^-_n,
\end{eqnarray}
sometimes termed the "classical" and "quantum" field variables. The two variables $\phi_0^q$ and $\phi_1^q$ are not sampled but integrated out directly in the path integral, which reduces the contour variables as shown in Figure \ref{fig:reduced_contour}.

Following the method outlined in \cite{mou2019real}, initial conditions consistent with a non-interacting thermal density matrix may be integrated out of the path integral and instead sampled from distributions given by 
\begin{align}
\label{eqn:initial_conditions}
    \left \langle \phi_0^{cl}(p) \left( \phi_0^{cl}(p')\right)^\dagger \right \rangle &= \frac{\hbar}{\omega_p}\left( n_p + \frac{1}{2}\right) (2\pi)^d\delta^d(p-p'), \nonumber \\
    \left \langle \dot{\phi}_0^{cl}(p) \left( \dot{\phi}_0^{cl}(p')\right)^\dagger \right \rangle &= \omega_p \hbar\left( n_p + \frac{1}{2}\right) (2\pi)^d\delta^d(p-p').
\end{align}

The initial correlators in Eq. (\ref{eqn:initial_conditions}) are defined in momentum space, and the objects $n_p$ and $\omega_p$ are the particle number and mode energy (or dispersion relation), respectively. In thermal equilibrium, a free scalar field would have a Bose-Einstein distribution and a standard relativistic dispersion relation, $\omega_p^2=p^2+m^2$. But in principle, one may choose anything, and thereby define some Gaussian non-equilibrium initial states. 

As we will discuss further below, we have now separated the Monte-Carlo sampling of the complete real-time path integral into two parts. First, the initial conditions $\phi_0^{cl}$ and $\phi_1^{cl}$ are sampled by drawing real values from a Gaussian distribution Eq. (\ref{eqn:initial_conditions}). For each such initial condition, we subsequently perform Monte-Carlo simulations using the Generalised Thimble method as described above, for all the remaining field variables $\phi_{n>1}^{cl}$, $\phi_{n>1}^q$, but keeping $\phi_0^{cl}$ and $\phi_1^{cl}$ fixed. Each initial condition uniquely determines a classical solution/critical point. At the same time, scanning over all initial conditions ensures that we include all critical points/thimbles in the system. 

%%%%%%%%%%%%%%%%%%%%%%%%%%%%%%%%%%%%%%%%%%%%%%%%%%%%%

\section{Numerical Simulations and Optimisations}
\label{sec:Numerics}

%%%%%%%%%%%%%%%%%%%%%%%%%%%%%%%%%%%%%%%%%%%%%%%%%%%%%

We have now set up a formalism to compute any real-time correlator exactly, up to the lattice discretization and the statistical error of the Monte-Carlo sampling. The sign problem is alleviated for finite $\tau$ and in principle resolved for $\tau\rightarrow\infty$, although this limit may not be reached in practice. 

In the following, we will investigate the scope and limitations of the formalism, from the point of view of one and two scalar fields. Spatial extent is at this stage less essential than time-extent, and we will proceed in 0+1 dimensions. We will comment briefly on 1+1 and 3+1 dimensional simulations in the Conclusions.

We will in the present section consider a simple scalar field model, for which the continuum action is
\begin{align}
S = \int d^Dx \left[\frac{1}{2}(\partial_\mu\phi)^2-\frac{m_\phi^2}{2}\phi^2\right].
\end{align}
In section \ref{sec:Two_fields}, this will be extended into a model of two interacting scalar fields, by adding
\begin{align}
+\int d^Dx \left[\frac{1}{2}(\partial_\mu\chi)^2-\frac{m_\chi^2}{2}\chi^2-\frac{\lambda_1}{4}\phi\chi-\frac{\lambda_2}{4}\phi^2\chi^2\right].
\end{align}
Whereas the first interaction term (proportional to $\lambda_1$) is really a non-diagonal mass contribution, leading only to mixing, the second one (proportional to $\lambda_2$) is a true non-linear interaction leading to particle decay. 

%%%%%%%%%%%%%

\subsection{Algorithm implementation}
\label{sec:Algorithm}

%%%%%%%%%%%%%%%

\begin{figure}
    \centering
    \includegraphics[width = 0.75\textwidth]{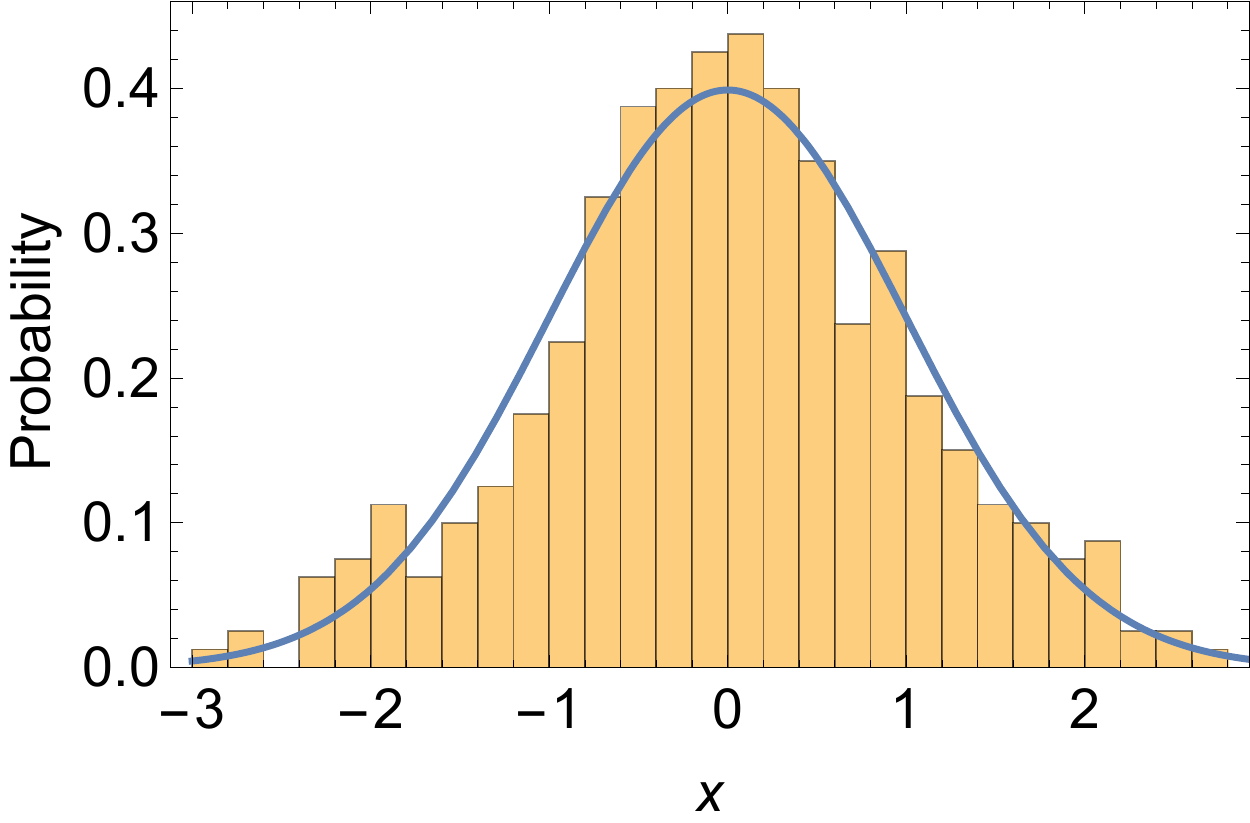}
    \caption{Histogram of 400 random numbers, sampling the Gaussian initial conditions.}
    \label{fig:Gaussian_sampling}
\end{figure}
We first set out the algorithm used to generate samples to compute the observables in Eq. (\ref{eqn:observable_samples}) \cite{mou2019real,Alexandru:2016gsd,Alexandru:2017lqr}. 

\begin{enumerate}
\item We assume that a discretized action for one or more scalar fields is given, involving a set of physical parameters (masses, couplings, ...). This action also involves non-physical parameters, such as the lattice spacings in time and space, the Keldysh contour time extent, and the chosen finite number of lattice points, in space and time.
\item We pick a maximum flow time $\tau_{max}$ and a flow time step $d\tau$. We also select a MC proposal width $\delta$. 
\item We define an initial condition through selecting the particle numbers $n_p$ and the mode energies $\omega_p$, for each lattice momentum mode. This may or may not be a thermal or vacuum state. 
\item As described above, we draw a set of $N_{init}$ values for the initial field variables $\phi_0^{cl}$, $\phi_1^{cl}$. Figure \ref{fig:Gaussian_sampling} shows an example of 400 initial values of $\phi_0^{cl}$ for one of the data sets described below. 
\item For each of these initial conditions, we first solve the classical equation of motion for the entire time extent on the lattice. This gives us values $\phi^{cl}_j$ for all times $j$, making up an initial configuration $\Tilde{\phi}_j^{cl}$ from which to start our Monte-Carlo sampling. As we also discussed above, the classical solution $\Tilde{\phi}_j^{cl}$ is a fixed point of the thimble gradient flow, a critical "point" in the multi-dimensional space spanned by all the $n$ complex planes. 
\item Now we construct a MC chain of configurations through a Metropolis-like algorithm. Given a current real "n-th" configuration $\varphi_j^n$ (which initially is the classical solution, $\Tilde{\phi}_j^{cl}$), we flow the entire configuration using Eq. (\ref{eqn:flow}) until $\tau=\tau_{max}$. That gives a complex-valued configuration $\phi_j^n$ including both $cl$ and $q$ variables. We then construct the Jacobian $J$ allowing for the transformation between complex-valued $\phi_j^n$ and real-valued $\varphi_j^n$. 
\item We randomly generate a complex proposal vector $\eta$ for all variables in the configuration (although not $\phi_0^{cl}$, $\phi_1^{cl}$), by drawing real and imaginary parts from a Gaussian with width $\sigma=\sqrt{2}\delta$.
\item We transform this into a proposal vector $\Delta$ on the real axis using $\eta =J\Delta$. This defines a new configuration $\varphi_j^{n+1}=\varphi_j^n+Re(\Delta)$.
\item We flow $\varphi_j^{n+1}$ to $\tau_{max}$ to give a proposal $\phi_j^{n+1}$.
\item We accept/reject the proposal with a probability \cite{mou2019real}
    \begin{eqnarray}
        Pr = \text{min}\left(e^{-(A_{n+1}-A_n)}, 1\right),
\end{eqnarray}
where 
\begin{eqnarray}
A_n=
\text{Re}(\mathcal{I}_n) - 2\ln|\det J_{n}|  + \Delta^T(J_n^\dagger J_n) \Delta/\delta^2  ,
    \end{eqnarray}
Note that this involves not just the difference in action but also the Jacobian. $\I_{n,n+1}$ and $J_{n,n+1}$ are implied to be evaluated at $\varphi_j^{n,n+1}$ respectively.
\item We repeat from step 6, until a sufficently long MC chain is generated, say $N_{MC}$ steps.
\item Then we start over from step 6 with a new initial classical configuration generated in steps 4 and 5.
\end{enumerate}
All the parameters $\tau_{max}$, $d\tau$, $\delta$, $N_{MC}$, $N_{init}$ may be optimised for best statistical significance and minimal numerical wall time. The optimal values may depend on the physical parameters in the action and the lattice size and discretization. 

%%%%%%%%%%%%%%%%%%%%%%%%%%%%%%%%%%

\subsection{Optimisation with a single field}
\label{sec:Optimisation}

%%%%%%%%%%%%%%%%%%%%%%%%%%%%%%%%%%

\begin{figure}
    \centering
    \includegraphics[scale = 0.7 ]{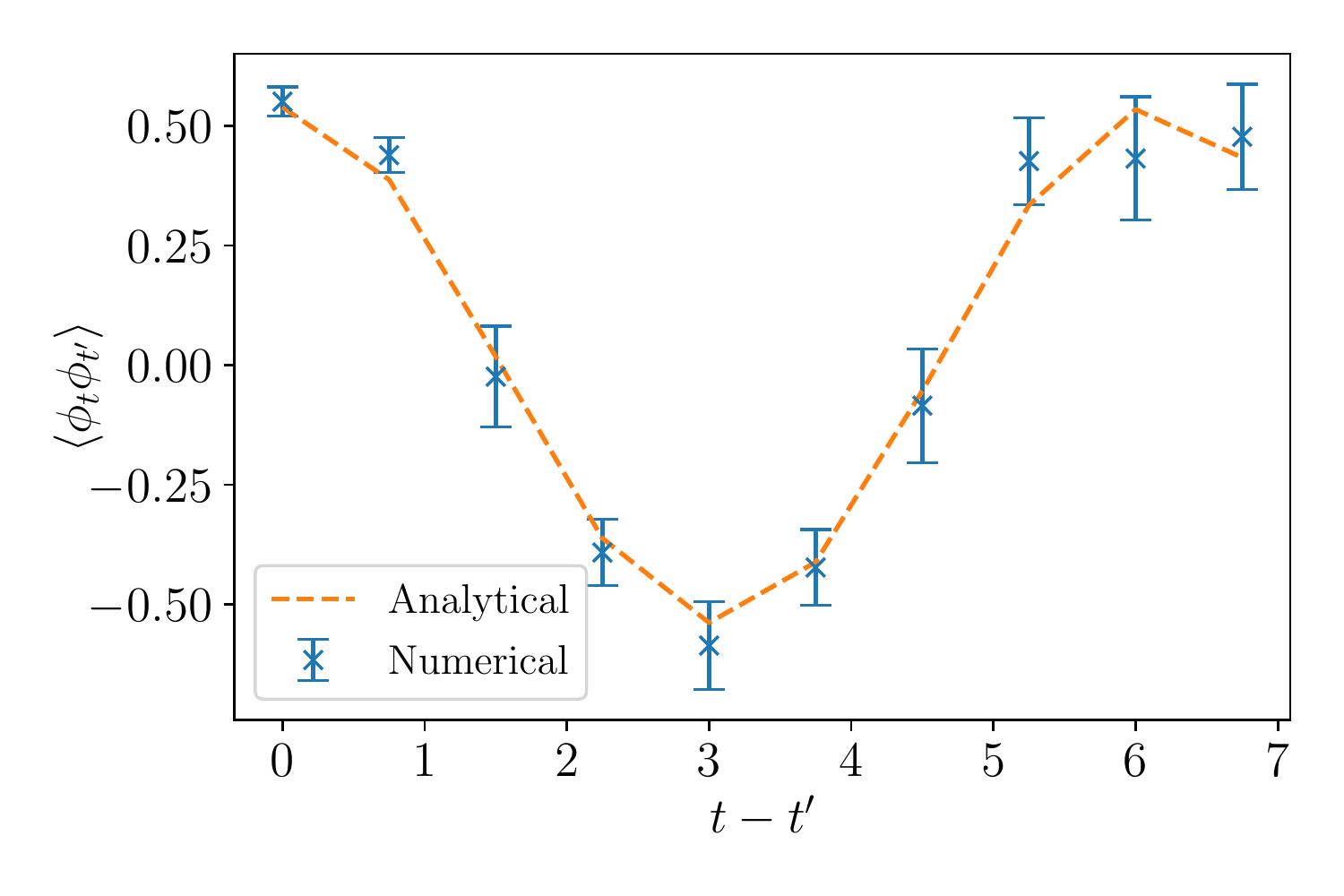}
   \caption{The correlator Eq. (\ref{eqn:correlator}) for a single free field.}
    \label{fig:correlator}
\end{figure}

We will perform our investigation of optimizing simulations with a single scalar field in 0+1 dimensions. We will consider one concrete test correlator, the non-equal time two-point function (or propagator) 
\begin{align}
\label{eqn:correlator}
\langle \phi_k^{cl} \phi_j^{cl} \rangle, 0 \leq k, j \leq m,
\end{align}
and gauge the performance based on how accurately we are able to determine this correlator.  An example is shown in Figure \ref{fig:correlator}.

After integrating out $\phi_{0,1}^q$, the discretized free field Lagrangian reads \cite{mou2019real}  (see Fig. \ref{fig:reduced_contour})
\begin{align}
    \mathcal{I} = &\left( \frac{-i}{\hbar}\right) \left[ \frac{2\phi_1 \Tilde{\phi}_2^{cl}}{\text{d}t} - \frac{\phi_2 \Tilde{\phi}_1^{cl}}{\td t} +  \frac{\phi_{2m - 2} \Tilde{\phi}_1^{cl}}{\td t} + \right.
    \left. \sum_{i = 1}^{2m -2} \frac{(\phi_{i + 1} - \phi_i)^2}{2\Delta_i} - \left( \frac{\Delta_i + \Delta_{i + 1}}{2}\right)\left( \frac{1}{2}m_\phi^2\phi_i^2\right)\right],
\end{align}
where 
\begin{equation}
    \Delta_i = \left\lbrace \begin{array}{cl}
         \td t, & 1 \leq i < m\\
         -\td t,& m \leq i < 2m - 1,
    \end{array}\right.
\end{equation}
The mass is taken to be $m_\phi=1$ in lattice units and $\hbar=1$ throughout.

The free field has a number of simplifying properties. Firstly, we know the exact solution for the two-point function to be a nicely oscillating (and therefore well-behaved) function, so that amplitudes and errors will be comparable for all $j,k$ pairs. Second, since the equation of motion is linear in $\phi$, the right-hand side of (\ref{eqn:Jacobian_calculation}) does not depend on $\phi_i$. As a result, the Jacobian in the flow evolution is constant, and does not need to be recomputed at every Monte Carlo and time step. This reduces the computational cost by about 95\%.

\begin{figure}
    \centering
    \begin{subfigure}[b]{0.49\textwidth}
        \centering
        \includegraphics[width = \textwidth]{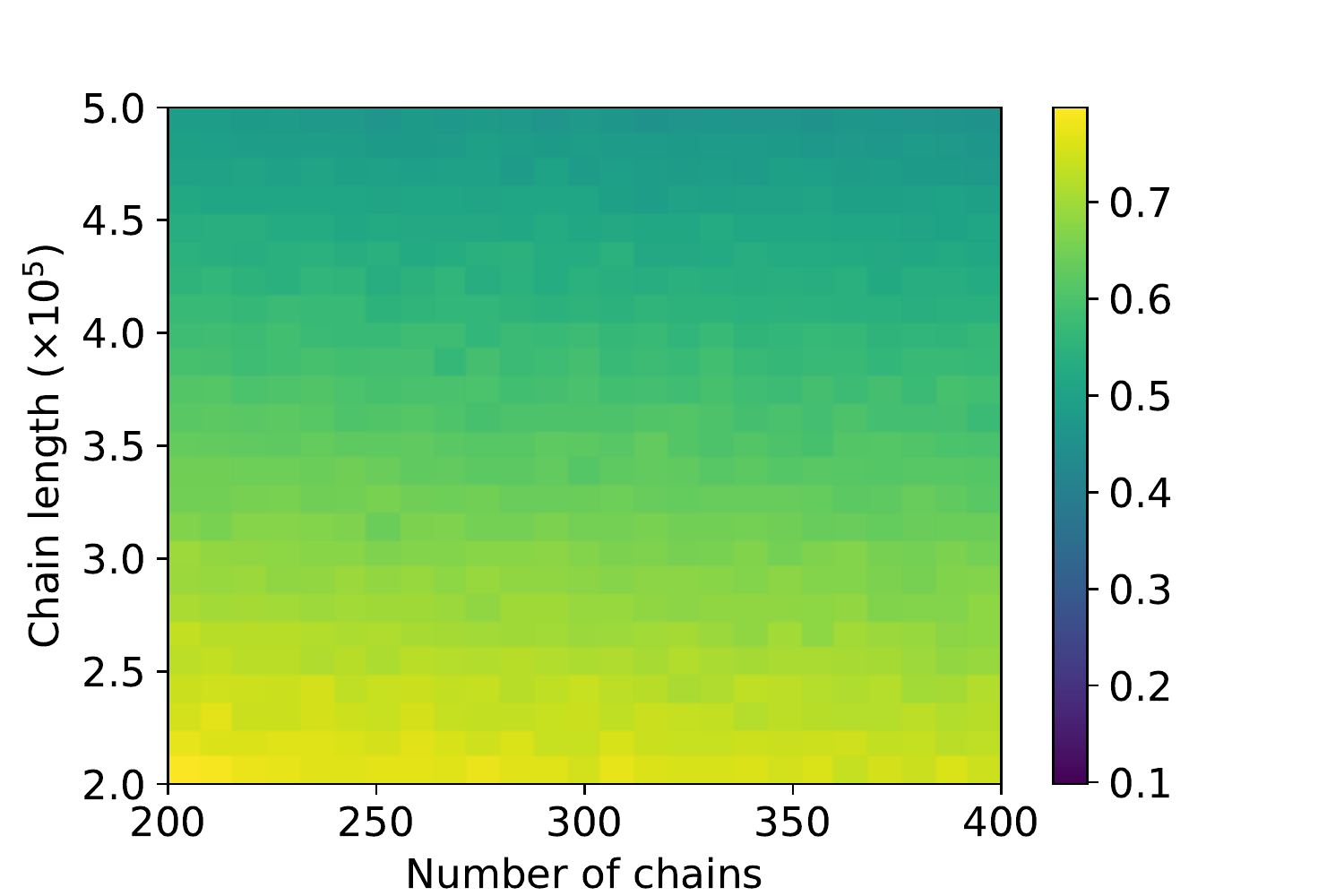}
        \caption{}
        \label{fig:cl_cn}
    \end{subfigure}
        \begin{subfigure}[b]{0.49\textwidth}
        \centering
        \includegraphics[width = \textwidth]{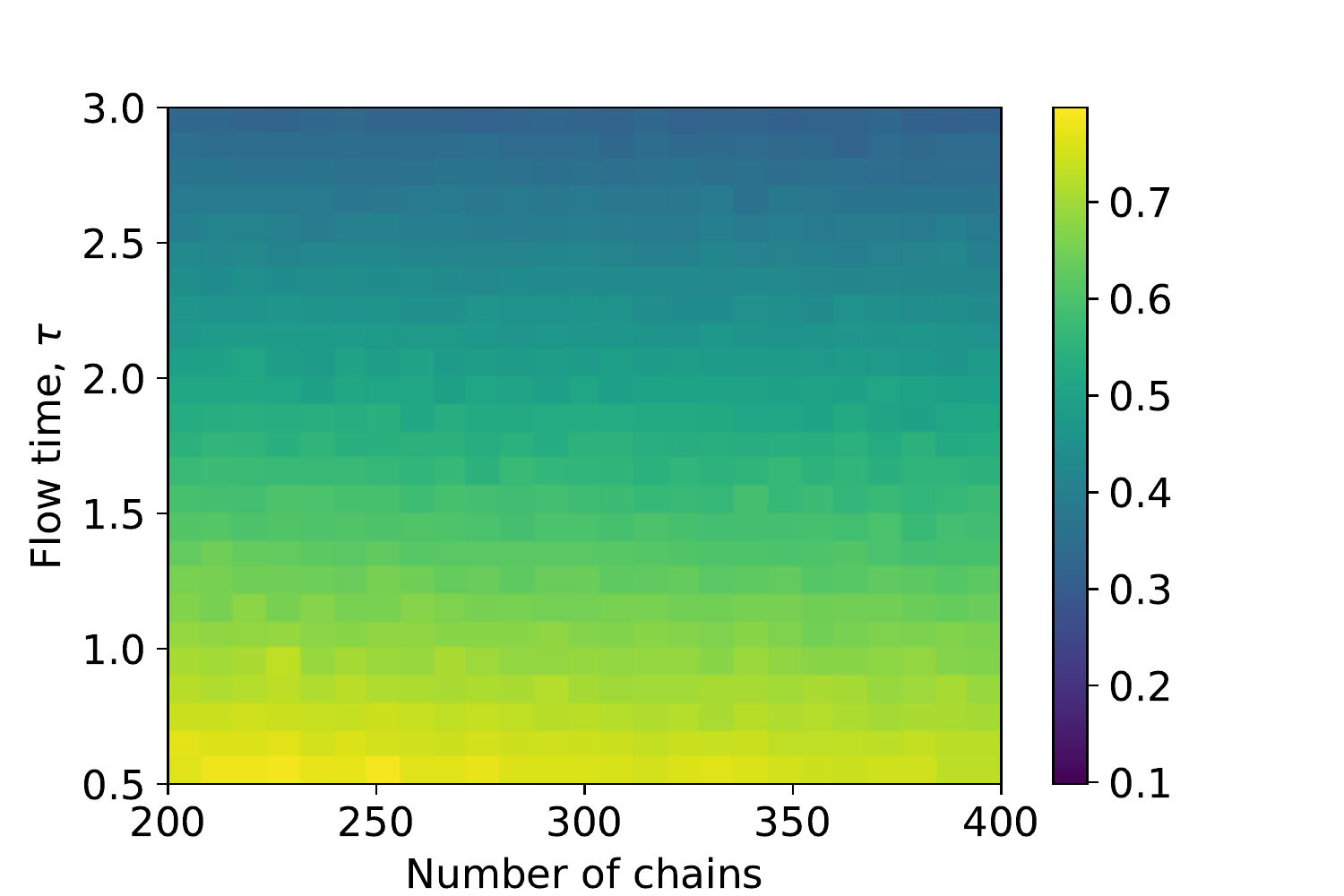}
        \caption{}
        \label{fig:cn_ft}
    \end{subfigure}
    \begin{subfigure}[b]{0.49\textwidth}
        \centering
        \includegraphics[width = \textwidth]{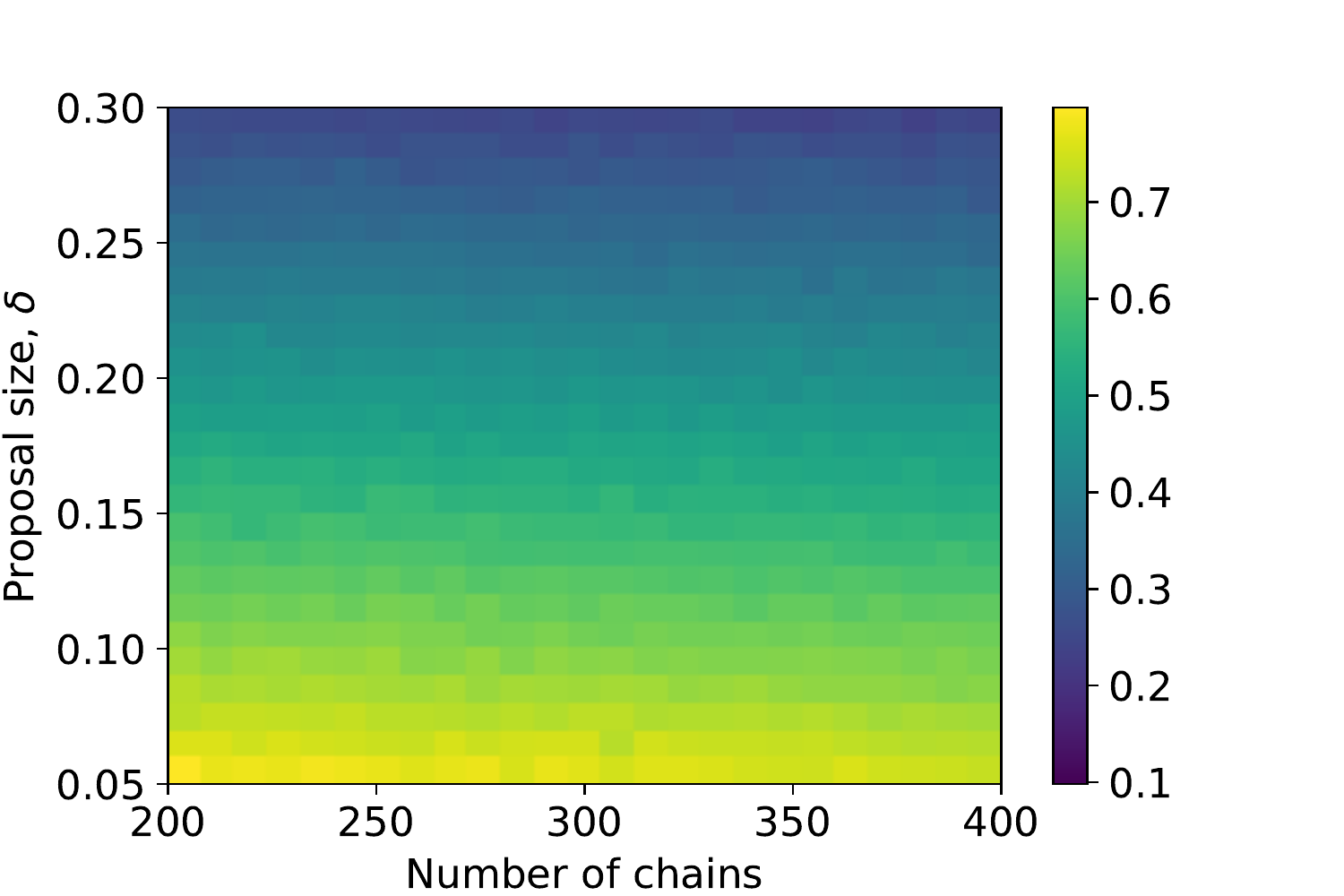}
        \caption{}
        \label{fig:cn_dl}
    \end{subfigure}
    \begin{subfigure}[b]{0.49\textwidth}
        \centering
        \includegraphics[width = \textwidth]{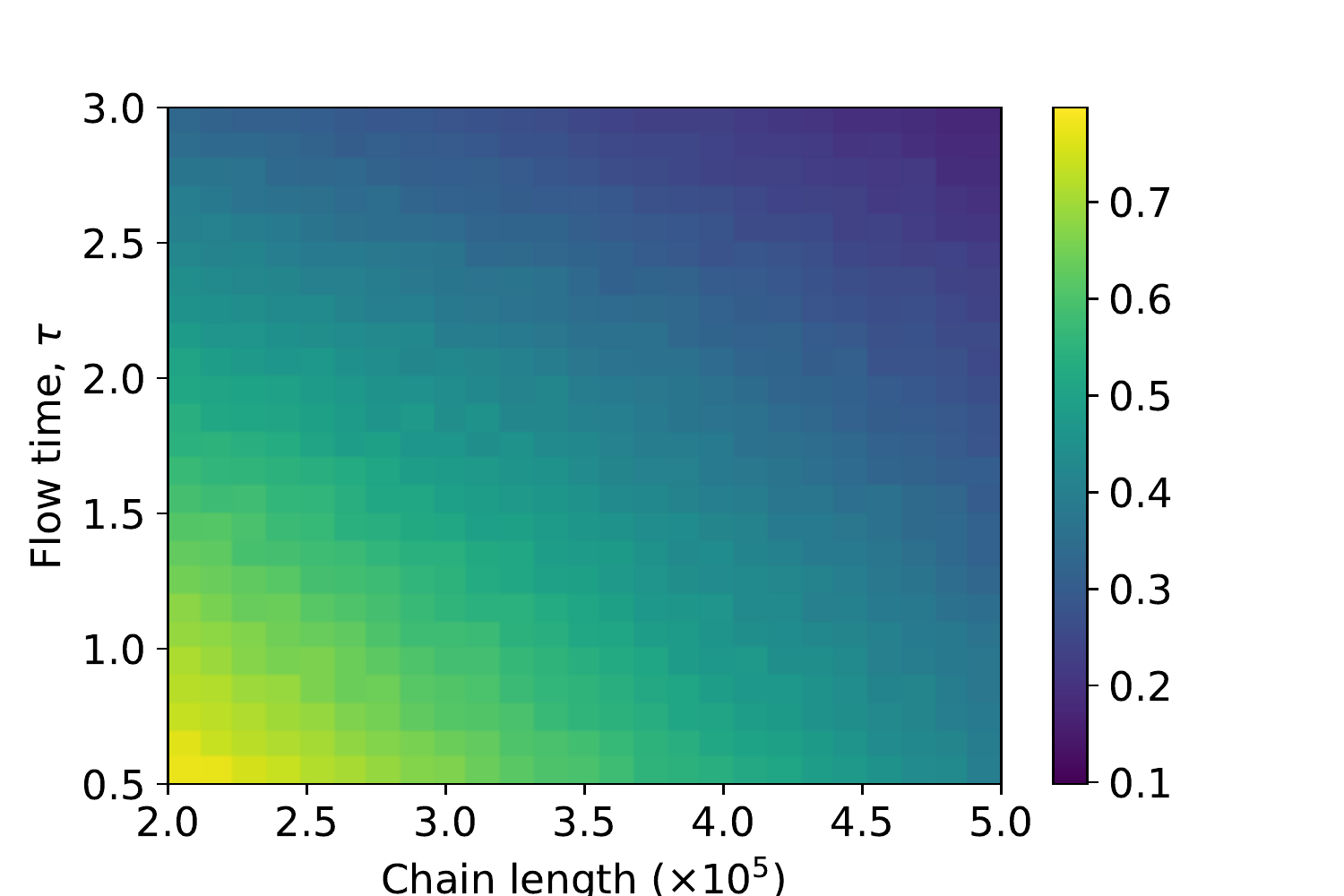}
        \caption{}
        \label{fig:cl_ft}
    \end{subfigure}
    \begin{subfigure}[b]{0.49\textwidth}
        \centering
        \includegraphics[width = \textwidth]{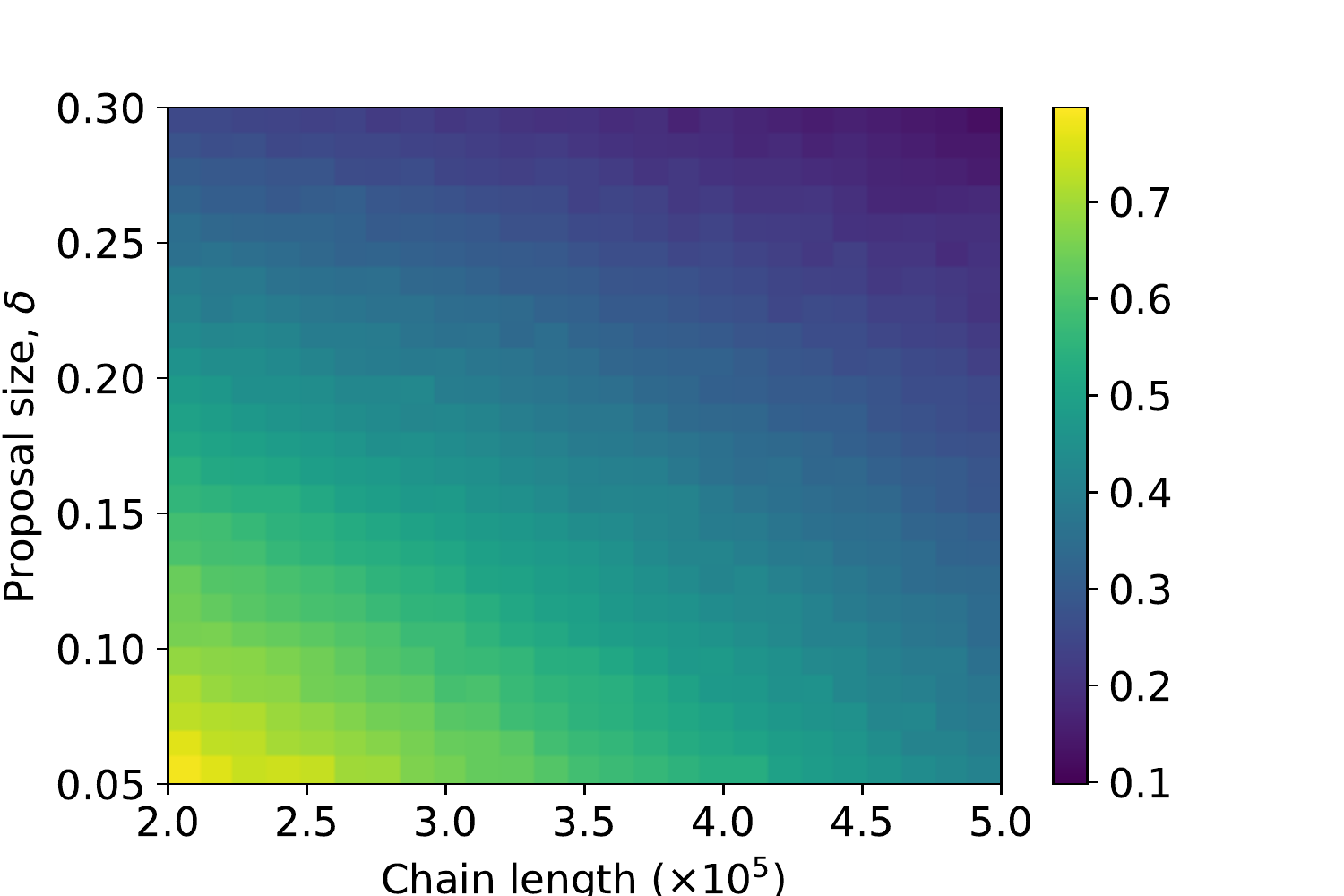}
        \caption{}
        \label{fig:cl_dl}
    \end{subfigure}
    \begin{subfigure}[b]{0.49\textwidth}
        \centering
        \includegraphics[width = \textwidth]{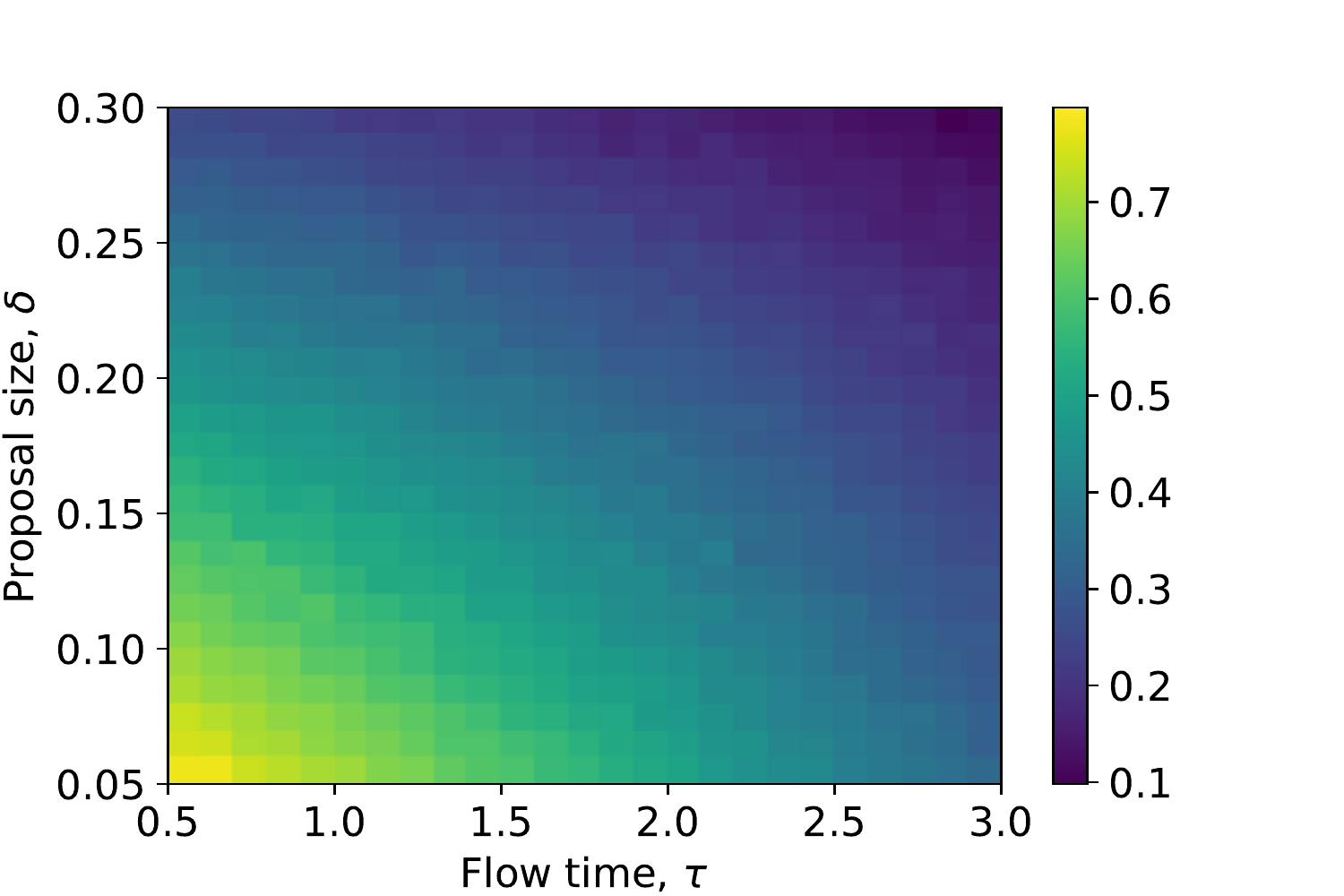}
        \caption{}
        \label{fig:ft_dl}
    \end{subfigure}
    \caption{A comparison of the maximum error on a test correlator for various simulation parameters}
    \label{fig:optimisation_plots}
\end{figure}

We select $dt=0.75$ and the number of time steps to be $m=10$. We then proceed to vary the number of initial conditions $N_{init}$ "number of MC chains", the flow time $\tau_{max}$, the parameter $\delta$ and the length of the MC chains $N_{MC}$. 

For the purpose of optimisation, we will define our "number of merit" to be the statistical error on the propagator, selecting the largest value over the $m=10$ time points. In Figure \ref{fig:optimisation_plots} we show correlation plots of this number of merit as we vary the parameters of the algorithm. We see that the performance improves with increasing MC chain length, increasing proposal size $\delta$ and increasing flow time $\tau$. The number of initial conditions is less important, provided it is large enough to convincingly sample the initial Gaussian distribution\footnote{For a purely classical simulation, the statistical error decreases as $N_{init}^{-1/2}$.} (see again Figure \ref{fig:Gaussian_sampling}).

\begin{figure}
    \centering
    \includegraphics[scale = 0.85 ]{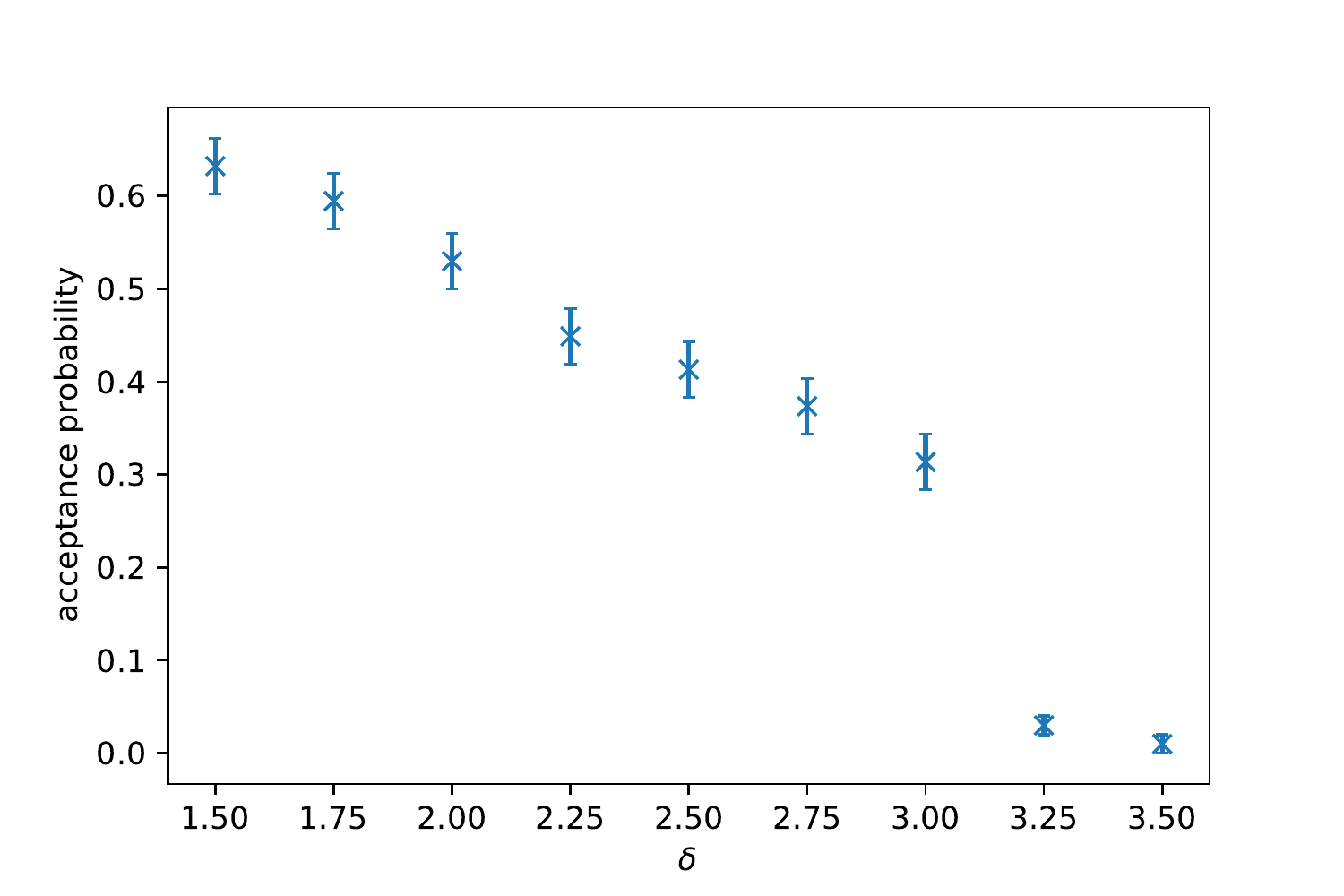}
    \caption{Probability of a proposal being accepted for $\tau = 1$. Note that while acceptance probability decreases with step size, the 'speed' around the manifold increases as the larger step size compensates.}
    \label{fig:acceptance_rate}
\end{figure}

We also see that the effects are uncorrelated, so that there is no favoured combination of parameters, that improves accuracy beyond the combined individual effects. The runtime depends linearly on $N_{init}$, $N_{MC}$ and $\tau_{max}$, since it is just how many times the algorithm is run. On the other hand, the runtime does not depend on $\delta$. This can however not be increased indefinitely, as shown in Figure \ref{fig:acceptance_rate}, which shows the acceptance rate of MC steps, as $\delta$ is increases. This drops substantially at a maximal value $\delta_{max}$ (in this case 3.15, for $\tau_{max}=1$). 

In general, the flowed field manifold can have a very complicated geometry. Having knowledge of the curvature in different directions along the manifold would allow us to generate random increments with different $\delta$ along each direction, for optimal speed through field configuration space. However, without such detailed knowledge of the geometry, we are left with selecting one, global $\delta$. The hope is that for a given set of parameters, we are able to identify $\delta_{max}$. Once the error can no longer be improved by increasing $\delta$, further improvements must come from increasing the flow time $\tau_{max}$ and the chain length $N_{MC}$. 

The effect of the chain length on the error is expected to be $\propto N_{MC}^{1/2}$, although there are considerations to do with the autocorrelation time. The dependence on the flow time seems to be approximately linear $\propto 1/\tau_{max}$.

So summarize: the flow time should be increased until the reward is cancelled by the corresponding $\delta_{max}$ decreasing. Once this has been optimized, any further computing power should be used to increase the chain length and the number of chains/initial conditions. The chain length should in any case at least be large enough that ergodicity is achieved and much longer than the autocorrelation time. Similarly, the number of chains/initial conditions must be large enough that the initial condition distribution is well sampled. Each chain is independent providing an excellent opportunity for parallelisation. 

\begin{figure}
    \centering
        \begin{subfigure}[b]{0.49\textwidth}
        \centering
        \includegraphics[width = \textwidth]{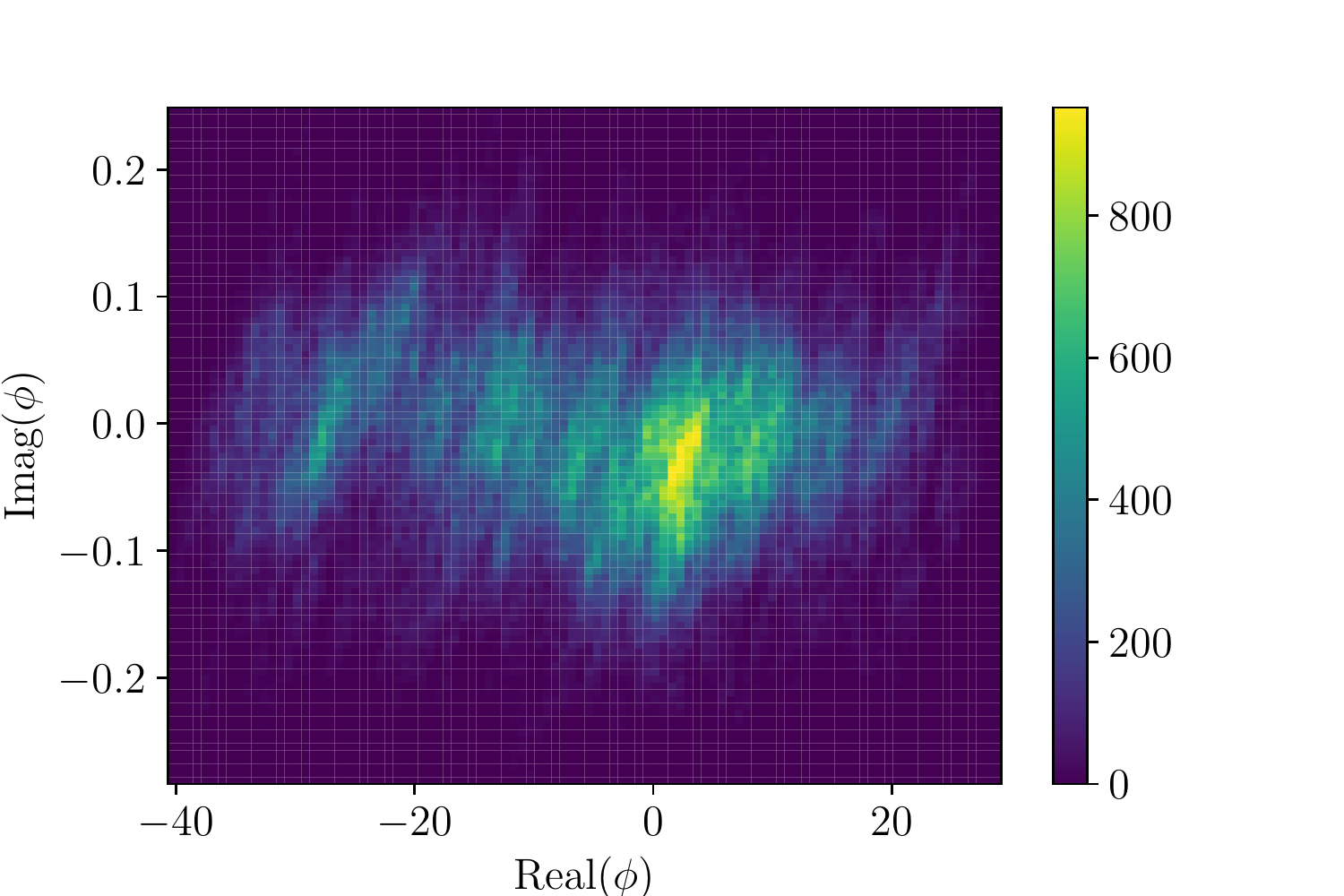}
        \caption{}
    \end{subfigure}
    \hfill
    \begin{subfigure}[b]{0.48\textwidth}
        \centering
        \includegraphics[width = \textwidth]{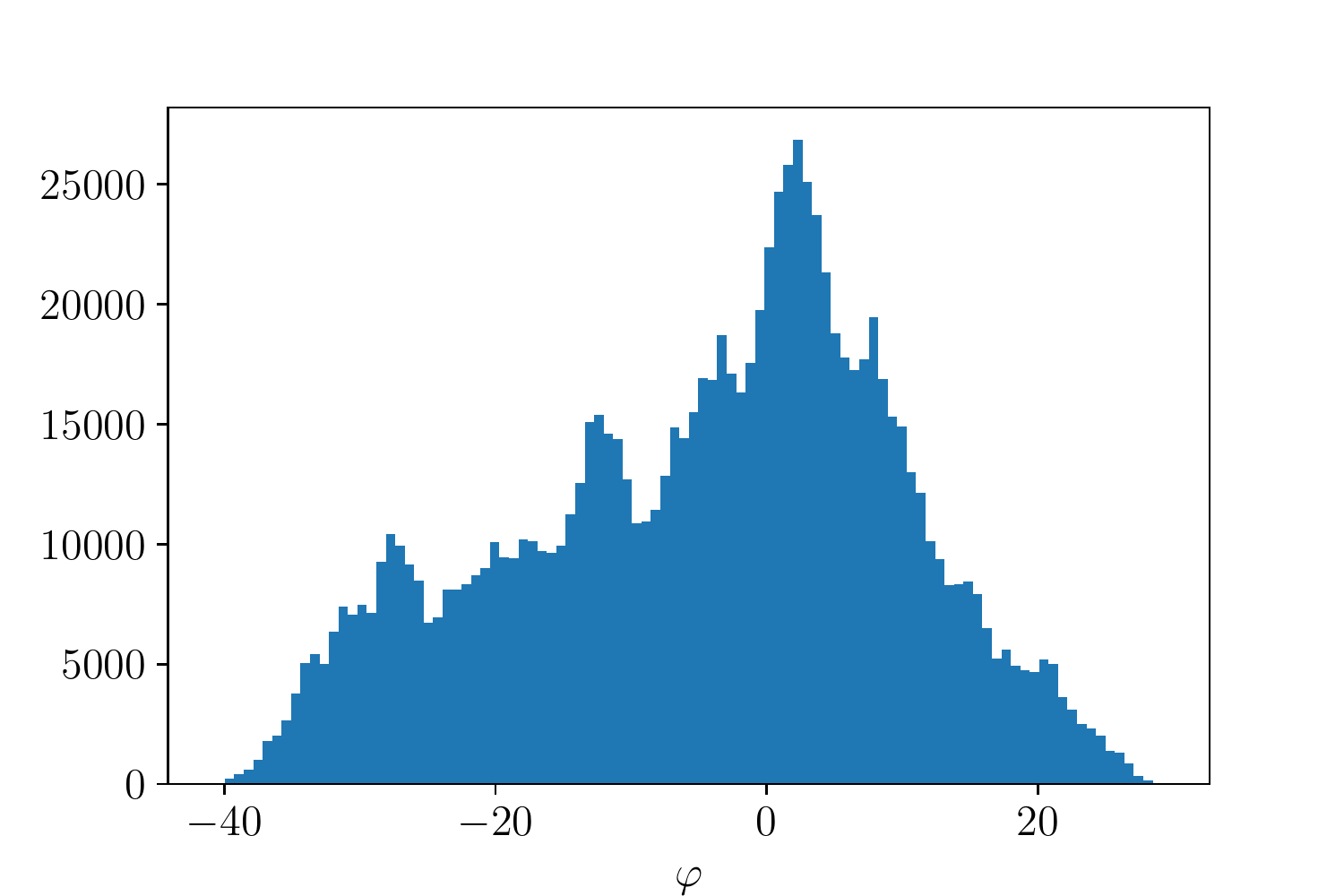}
        \caption {}
    \end{subfigure}
        \begin{subfigure}[b]{0.49\textwidth}
        \centering
        \includegraphics[width = \textwidth]{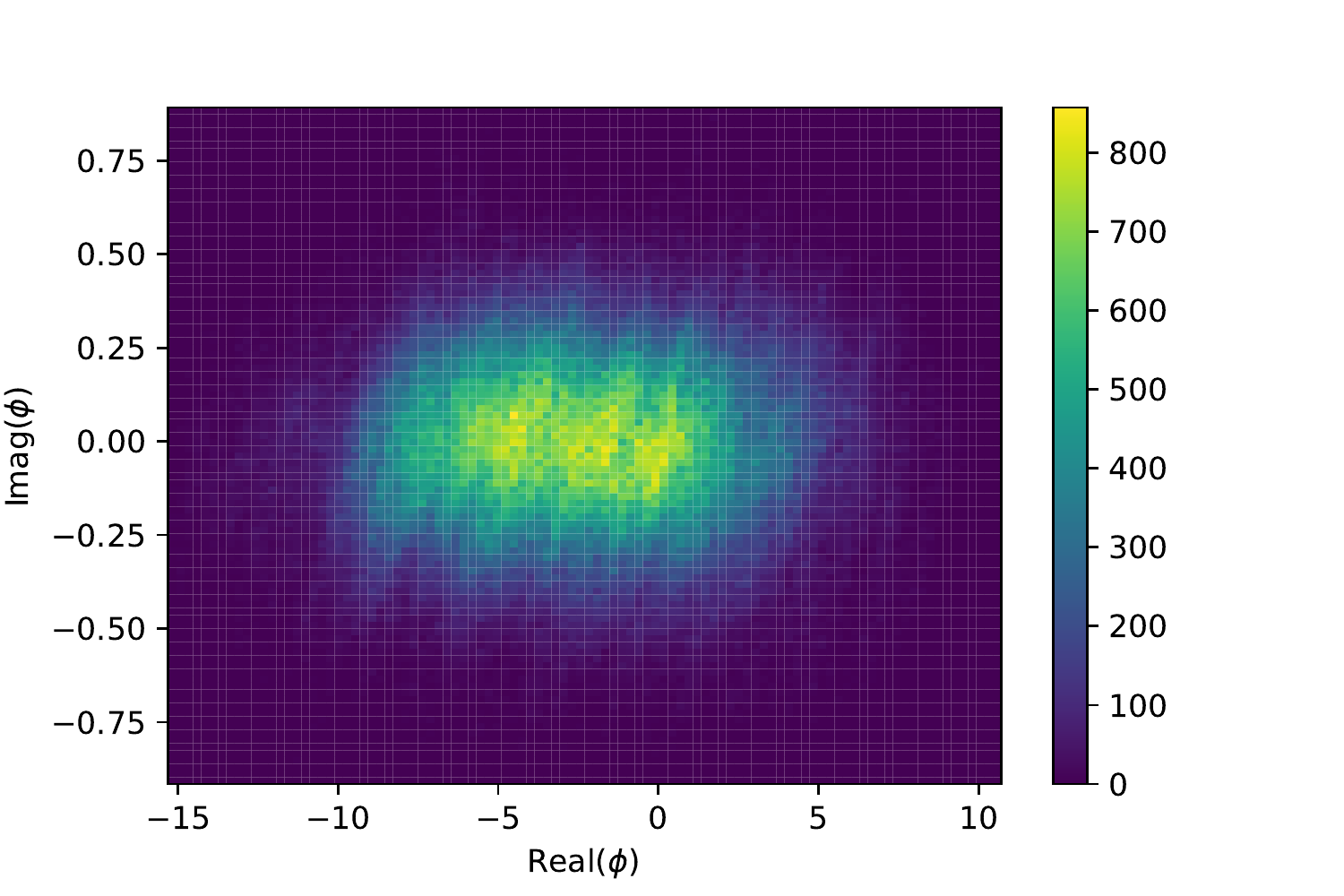}
        \caption{}
    \end{subfigure}
    \hfill
    \begin{subfigure}[b]{0.49\textwidth}
        \centering
        \includegraphics[width = \textwidth]{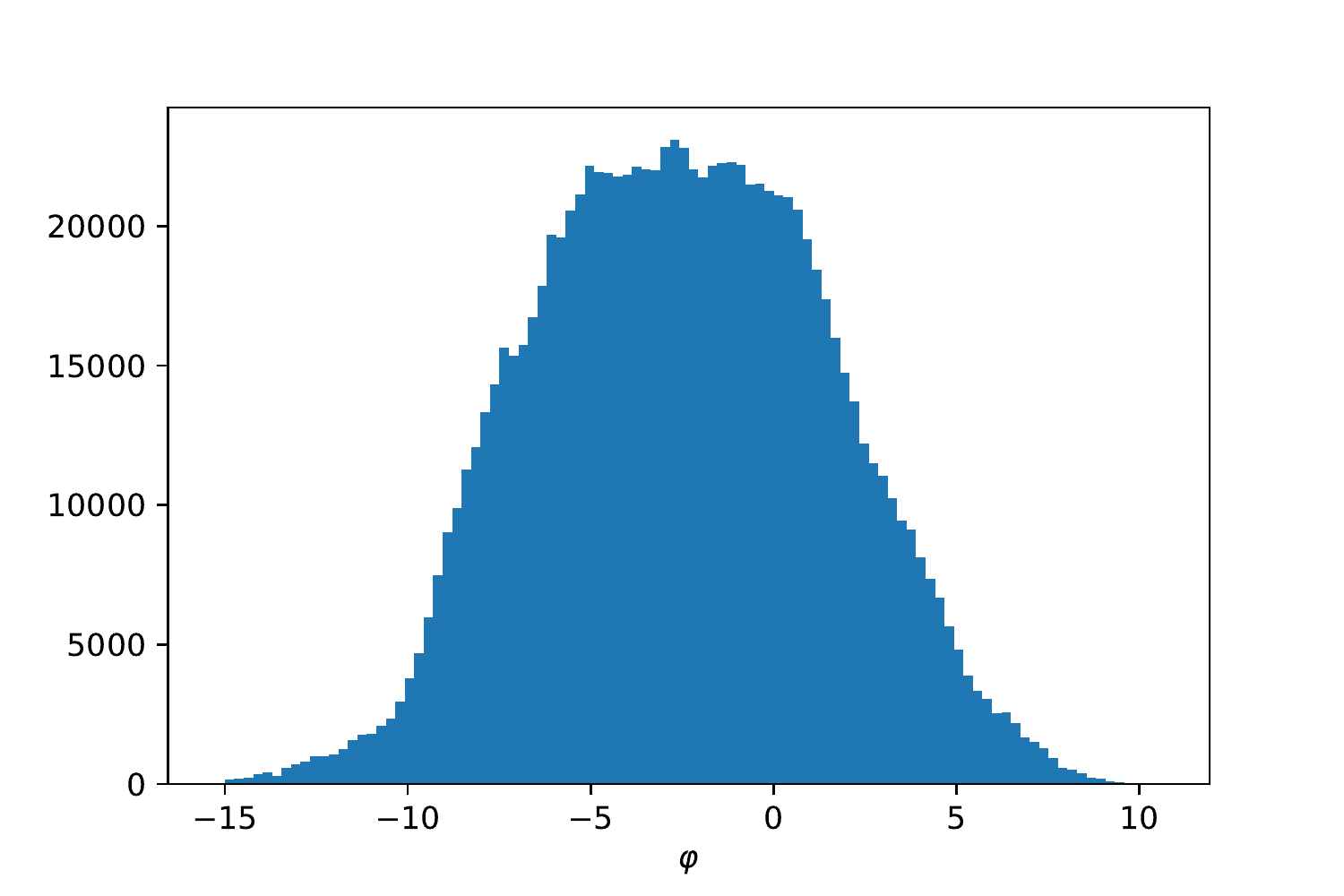}
        \caption {}
    \end{subfigure}
        \begin{subfigure}[b]{0.49\textwidth}
        \centering
        \includegraphics[width = \textwidth]{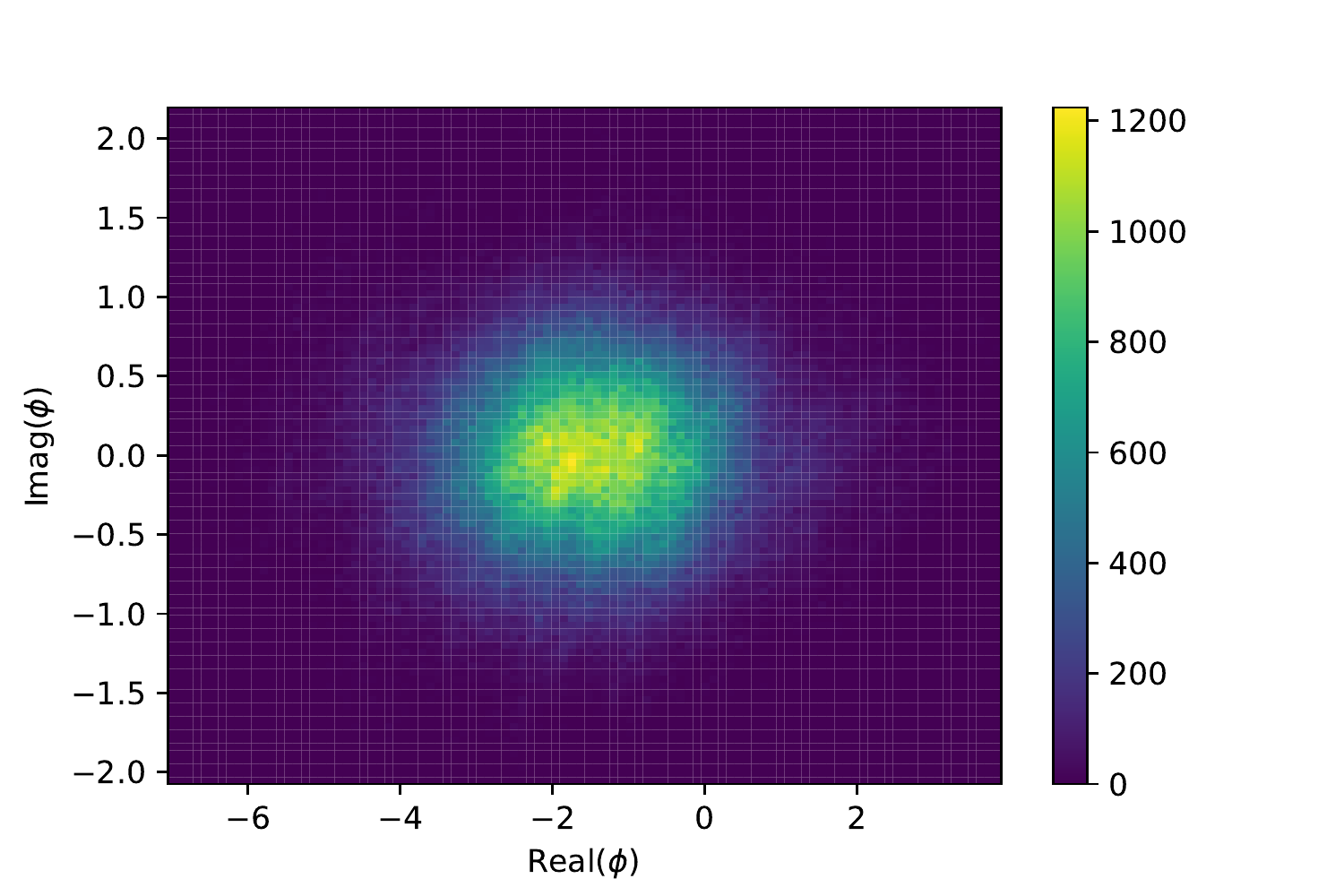}
        \caption{}
    \end{subfigure}
    \hfill
    \begin{subfigure}[b]{0.49\textwidth}
        \centering
        \includegraphics[width = \textwidth]{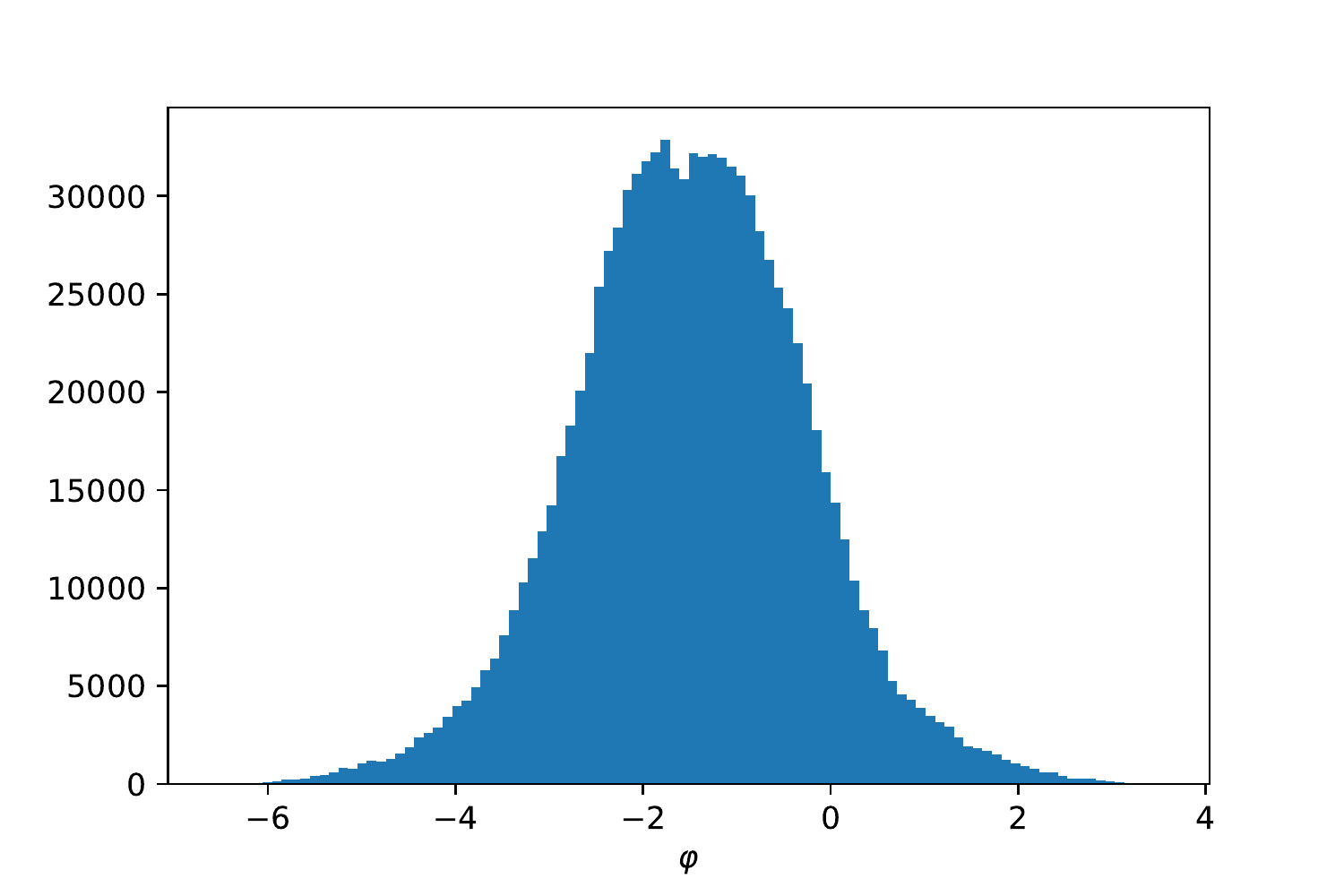}
        \caption {}
    \end{subfigure}
    \caption{Complex domain of a single field variable during the MC sampling of a multivariable system (left) and the corresponding domain in terms of the un-flowed real variable (right). Top to bottom, $\tau_{max}=0.01, 0.1, 1$.}
    \label{fig:onevar_thimble}
\end{figure}

For the simple one-variable example of Figure \ref{fig:airy_demo}, we were able to explicitly compute the Thimble and the Generalized Thimble, where the field manifold flows to. For multiple variables this is highly non-trivial, and in a MC sampling of coupled variables, it is the entire multidimensional manifold, including initial conditions, that is sampled. Still, it may be illustrative to show the domain in the complex plane, that one single variable samples during the course of the entire MC simulation. This will depend on the flow time $\tau_{max}$, where for $\tau_{max}=0$, the domain is the real axis. In Figure \ref{fig:onevar_thimble}, we show this domain for three different flow times, $\tau_{max}=0.01, 0.1, 1$. We see that for larger flow times, a larger region of the complex plane is sampled (note the different scales on the axes). In the right-hand panels, we see the corresponding distribution of the real-valued variables $\varphi$. As the flow time becomes larger, they cluster around an ever smaller range near, but displaced from, the origin. This is qualitatively similar to the one-variable example.

%%%%%%%%%%%%%%%%%%%%%%%%

\section{An interacting two-field system}
\label{sec:Two_fields}

%%%%%%%%%%%%%%%%%%%%%%%%

\begin{figure}
    \centering
    \begin{subfigure}[b]{0.49\textwidth}
        \centering
        \includegraphics[width = \textwidth]{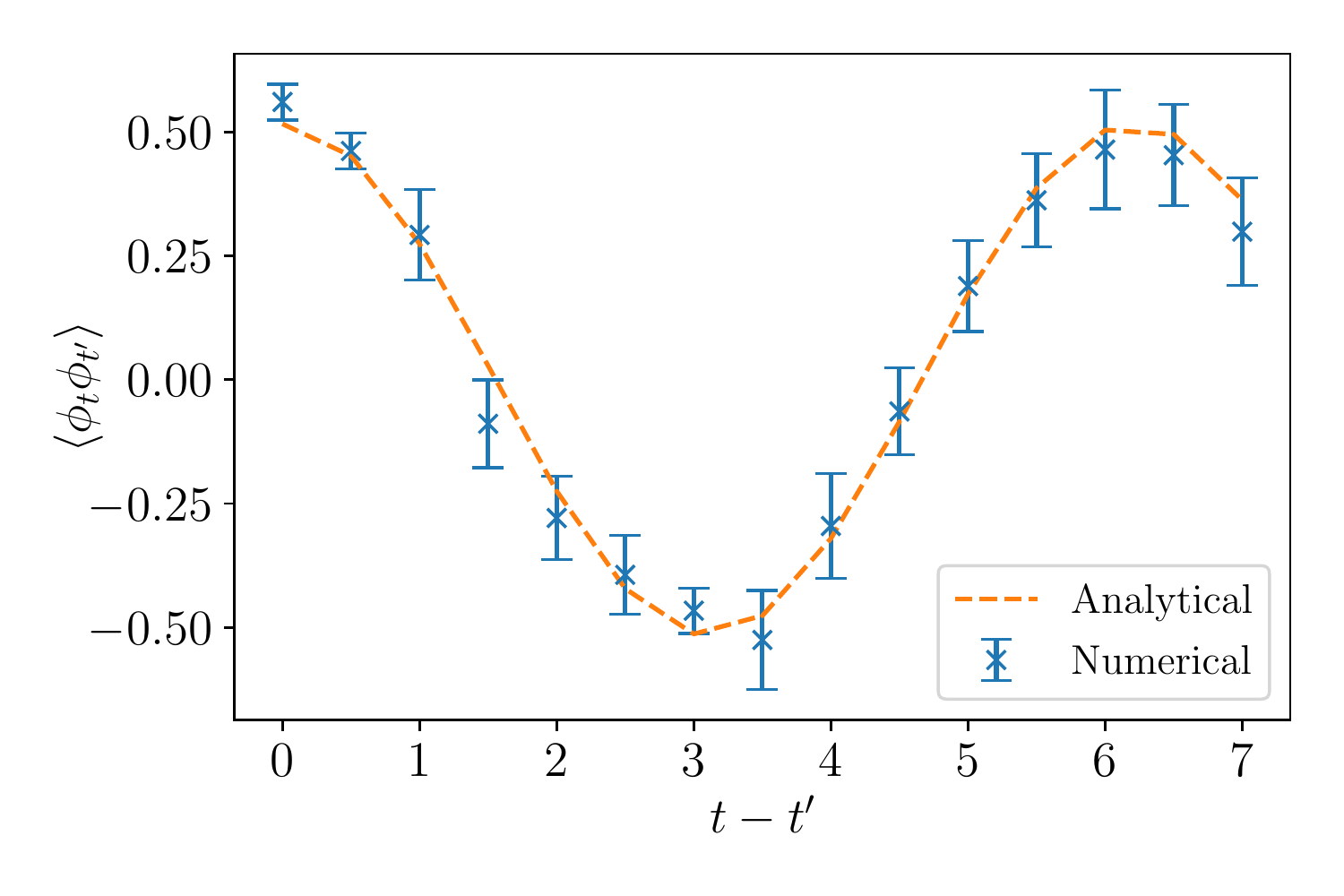}
        \caption{Free field $\langle \phi \phi \rangle$ un-equal time correlator}
        \label{fig:phi_analytic_comparison}
    \end{subfigure}
    \hfill
    \begin{subfigure}[b]{0.49\textwidth}
        \centering
        \includegraphics[width = \textwidth]{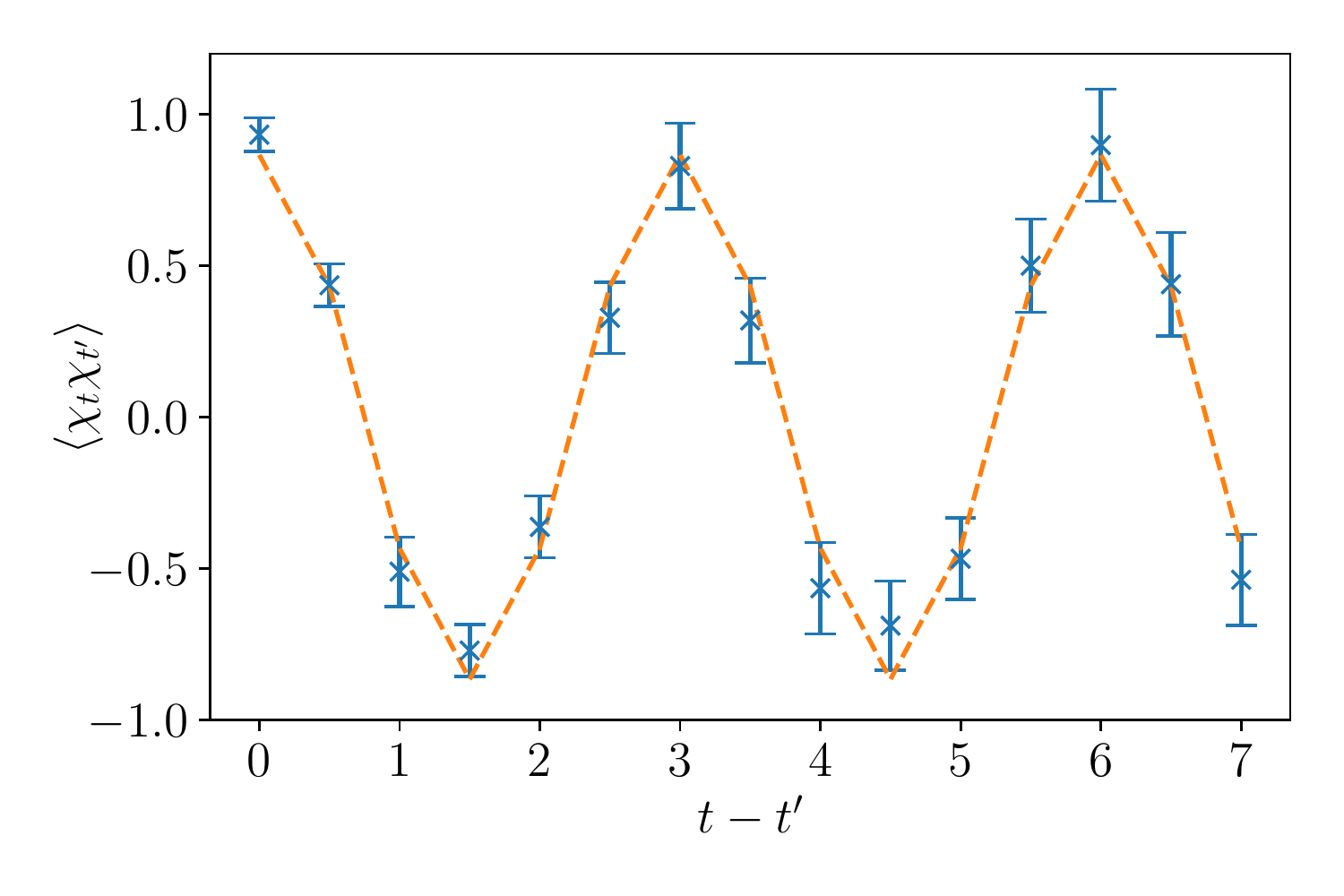}
        \caption{Free field $\langle \chi \chi \rangle$ un-equal time correlator}
        \label{fig:chi_analytic_comparison}
    \end{subfigure}
\newline
\vspace{0.1cm}

\begin{subfigure}[b]{0.49\textwidth}
        \centering
        \includegraphics[width = \textwidth]{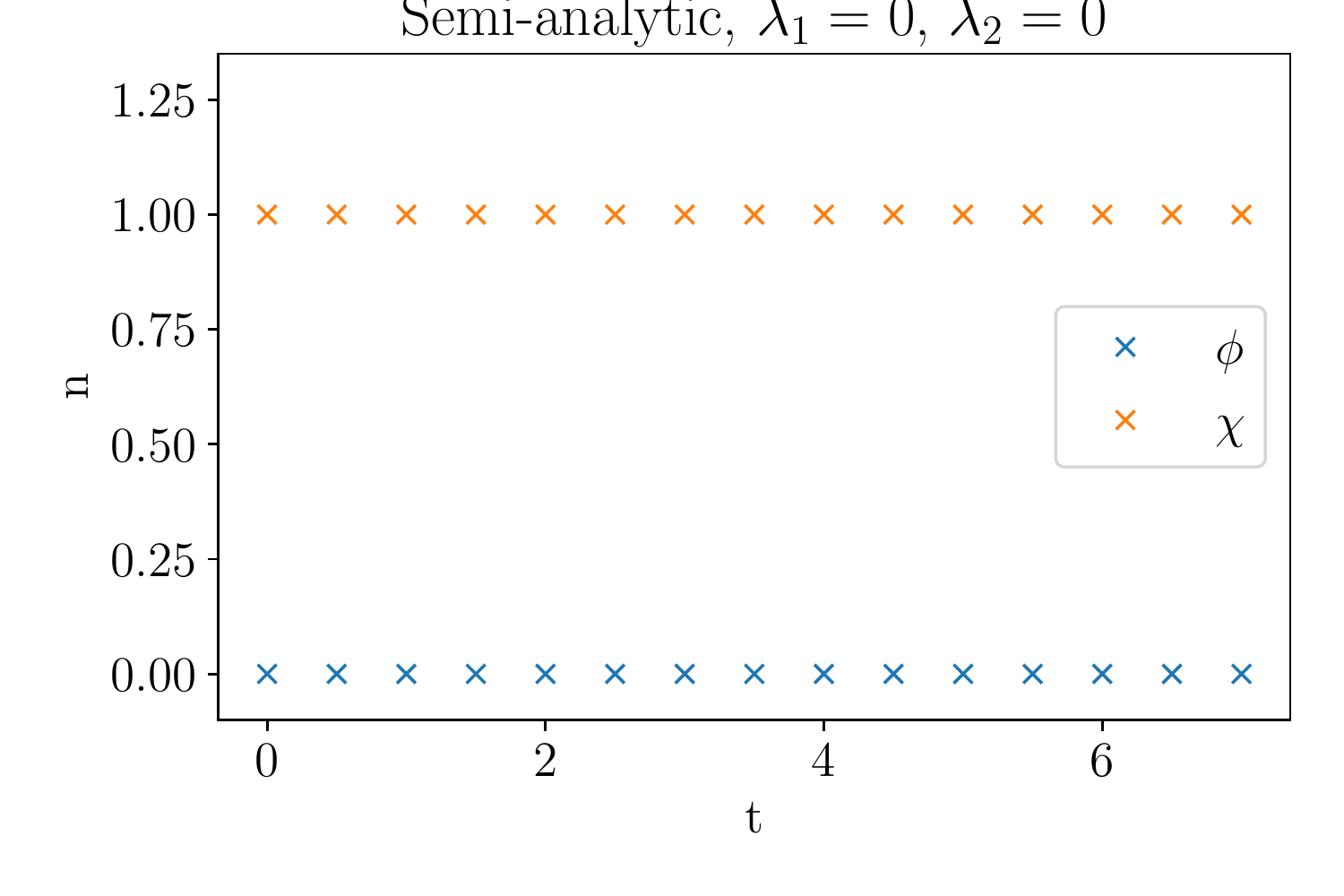}
        \caption{Free field $\phi$ and $\chi$ occupation numbers, analytic computation.}
    \end{subfigure}
    \hfill
    \begin{subfigure}[b]{0.49\textwidth}
        \centering
        \includegraphics[width = \textwidth]{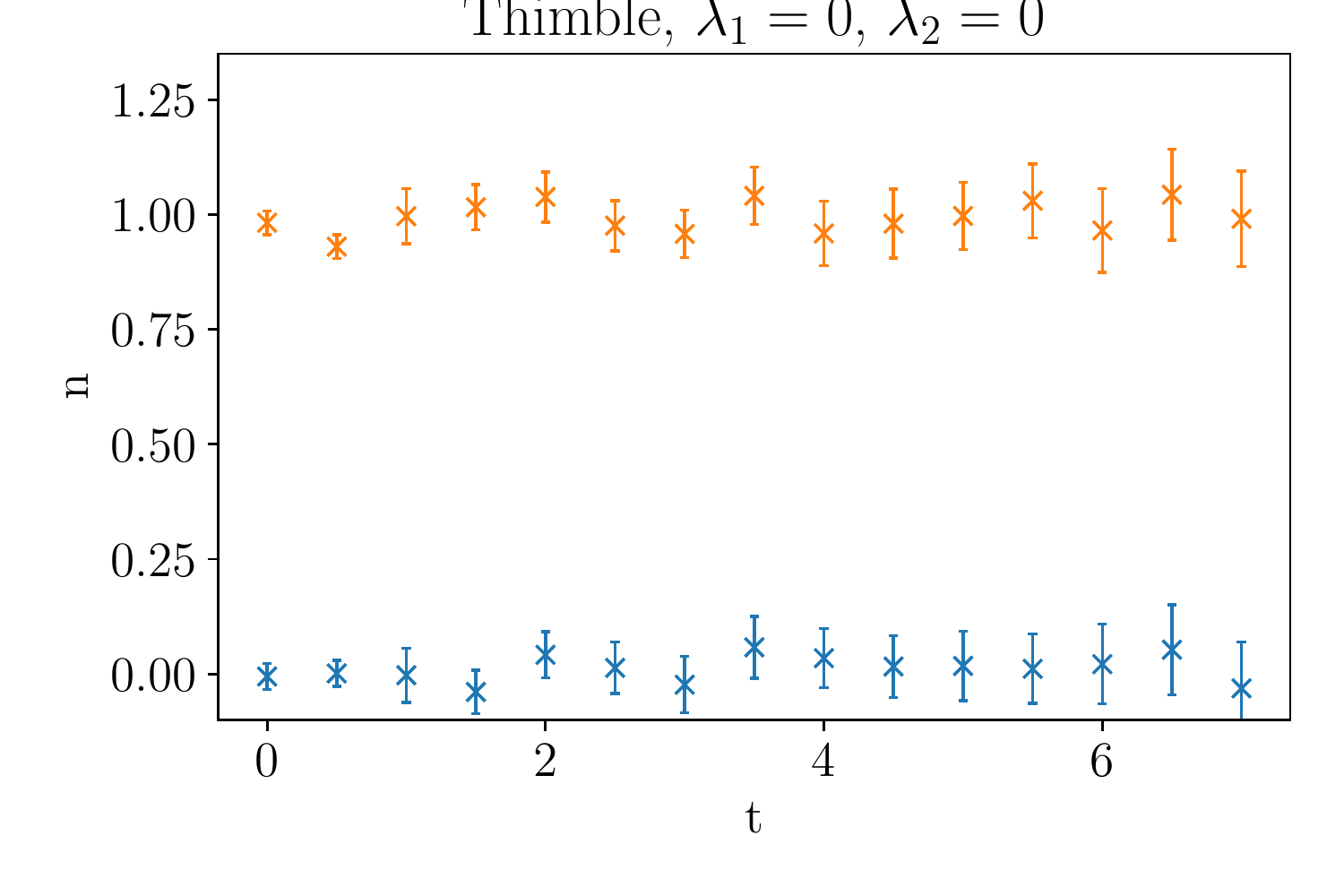}
        \caption {Free field $\phi$  and $\chi$ occupation numbers, Thimble computation.}
    \end{subfigure}
    \caption{The un-equal time correlator (top)  and occupation number (bottom) for two free fields, comparing semi-analytic results to Thimble results.}
    \label{fig:analytic_comparison}
\end{figure}

Using the optimisations outlined in section \ref{sec:Optimisation} allows us to add a second field to our simulations without exceeding our computational resources. This second field exists on the same lattice as our original field, and is implemented through the action 
\begin{align}
    \mathcal{I} = & \left( \frac{-i}{\hbar} \right)\left[ \frac{2\phi_1 \Tilde{\phi}_2^{cl}}{\td t} - \frac{\phi_2 \Tilde{\phi}_1^{cl}}{\td t} + \frac{\phi_{2m-2} \Tilde{\phi}_1^{cl}}{\td t} +\frac{2\chi_1 \Tilde{\chi}_2^{cl}}{\td t} - \frac{\chi_2 \Tilde{\chi}_1^{cl}}{\td t} + \frac{\chi_{2m-2} \Tilde{\chi}_1^{cl}}{\td t} \right. \nonumber
    \\ & \left. + \sum_{i = 1}^{2m - 2}\frac{(\phi_{i + 1} - \phi_i)^2}{2\Delta_i} + \frac{(\chi_{i + 1} - \chi_i)^2}{2\Delta_i} - \left( \frac{\Delta_i + \Delta_{i + 1}}{2}\right)\left( \frac{1}{2}m_\phi^2 \phi_i^2 + \frac{1}{2}m_\chi^2 \chi_i^2 + \frac{\lambda_1}{4}\phi_i \chi_i+\frac{\lambda_2}{4}\phi_i^2 \chi_i^2\right)\right],
\end{align}
where $\chi$ represents the second field. We have included a bilinear mass mixing term parameterized by $\lambda_1$ and a quartic interaction parameterized by $\lambda_2$. When $\lambda_1=\lambda_2=0$, we recover two decoupled free systems, as in the previous section. When $\lambda_1\neq 0$, the system is still free, but $\phi$ and $\chi$ are no longer mass eigenstates, and oscillation between the two states is expected. When $\lambda_2\neq 0$, we can expect actual interactions, decay and scattering between the two, depending on parameter values and the initial condition. 

We will focus on the case when $\chi$ is the heavier field, and initially occupied, and the $\phi$ is the lighter field and initially in vacuum. Concretely, we take $n_p = 0$ for $\phi$ and $n_p = 1$ for $\chi$, $m_\phi = 1$ and $m_\chi = 2$. Considering first $\lambda_1=\lambda_2=0$, we display the free correlators in Figure \ref{fig:analytic_comparison}. As the system is really two-variable quantum mechanics, we can in fact solve the system semi-analytically using the method described in appendix \ref{app:Heisenberg}, and use this for comparison. All thimble results below were generated with 400 chains of length $2 \times 10^6$, with $\tau = 1.5$, $\td t = 0.5$ and $\delta = 0.27$. 

%%%%%%%%%%%%%%%%%%%%%%%%%%%%%%%%%%

\subsection{Mass mixing and field oscillations}
\label{sec:mixing}

%%%%%%%%%%%%%%%%%%%%%%%%%%%%%%%%%%

\begin{figure}
    \centering
        \begin{subfigure}[b]{0.49\textwidth}
        \centering
        \includegraphics[width = \textwidth]{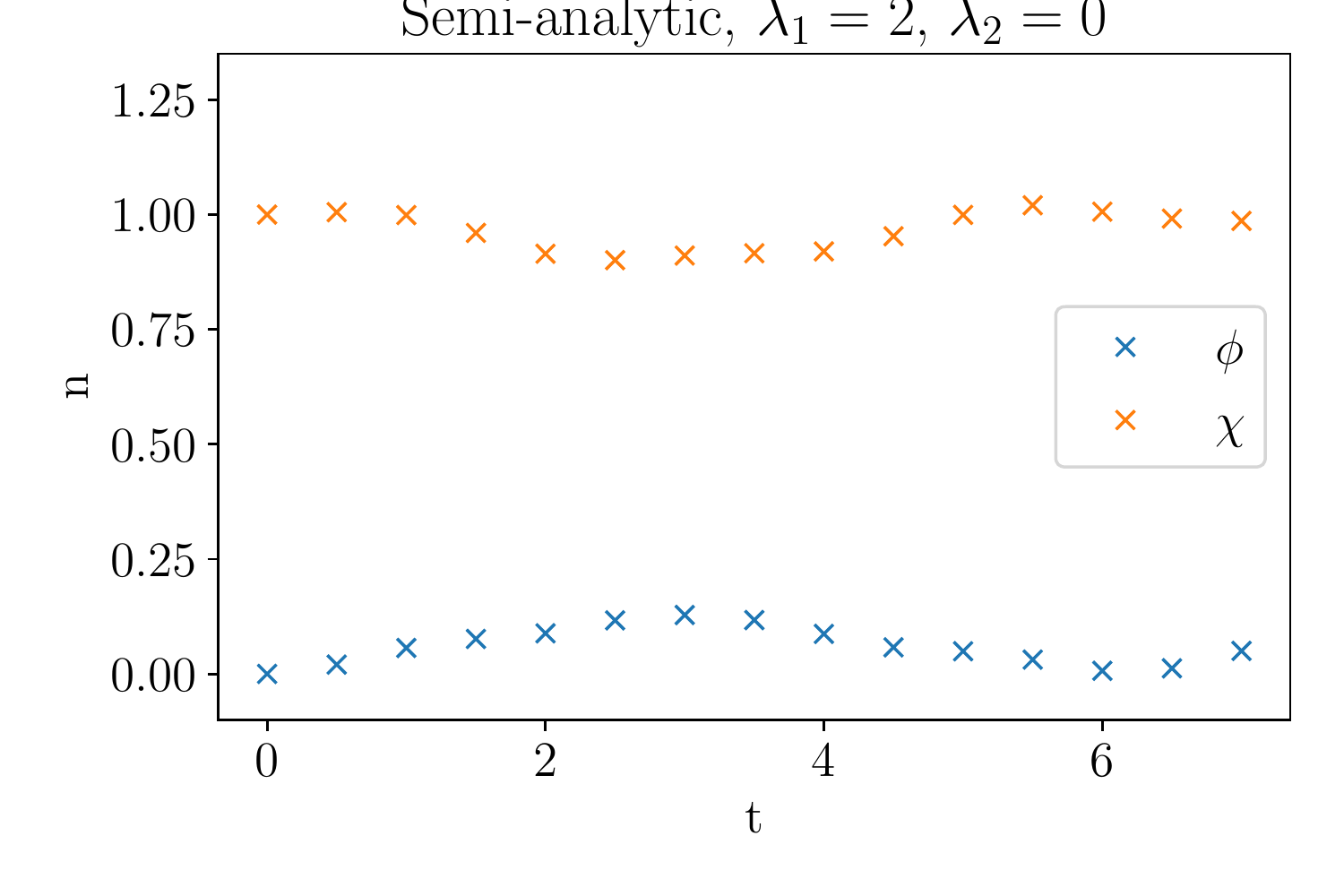}
        \caption{}
    \end{subfigure}
    \hfill
    \begin{subfigure}[b]{0.48\textwidth}
        \centering
        \includegraphics[width = \textwidth]{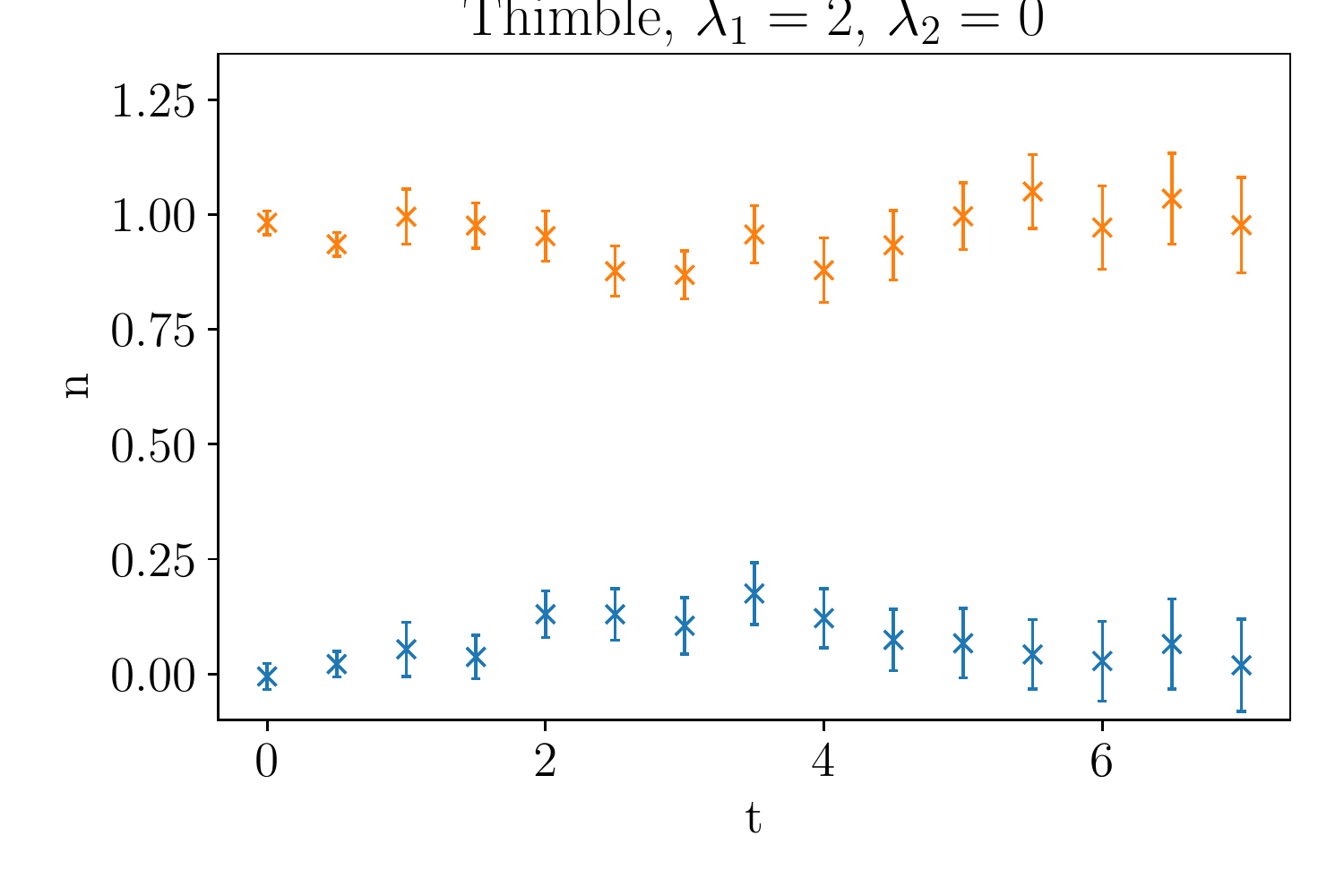}
        \caption {}
    \end{subfigure}

\vspace{0.2cm}

        \begin{subfigure}[b]{0.49\textwidth}
        \centering
        \includegraphics[width = \textwidth]{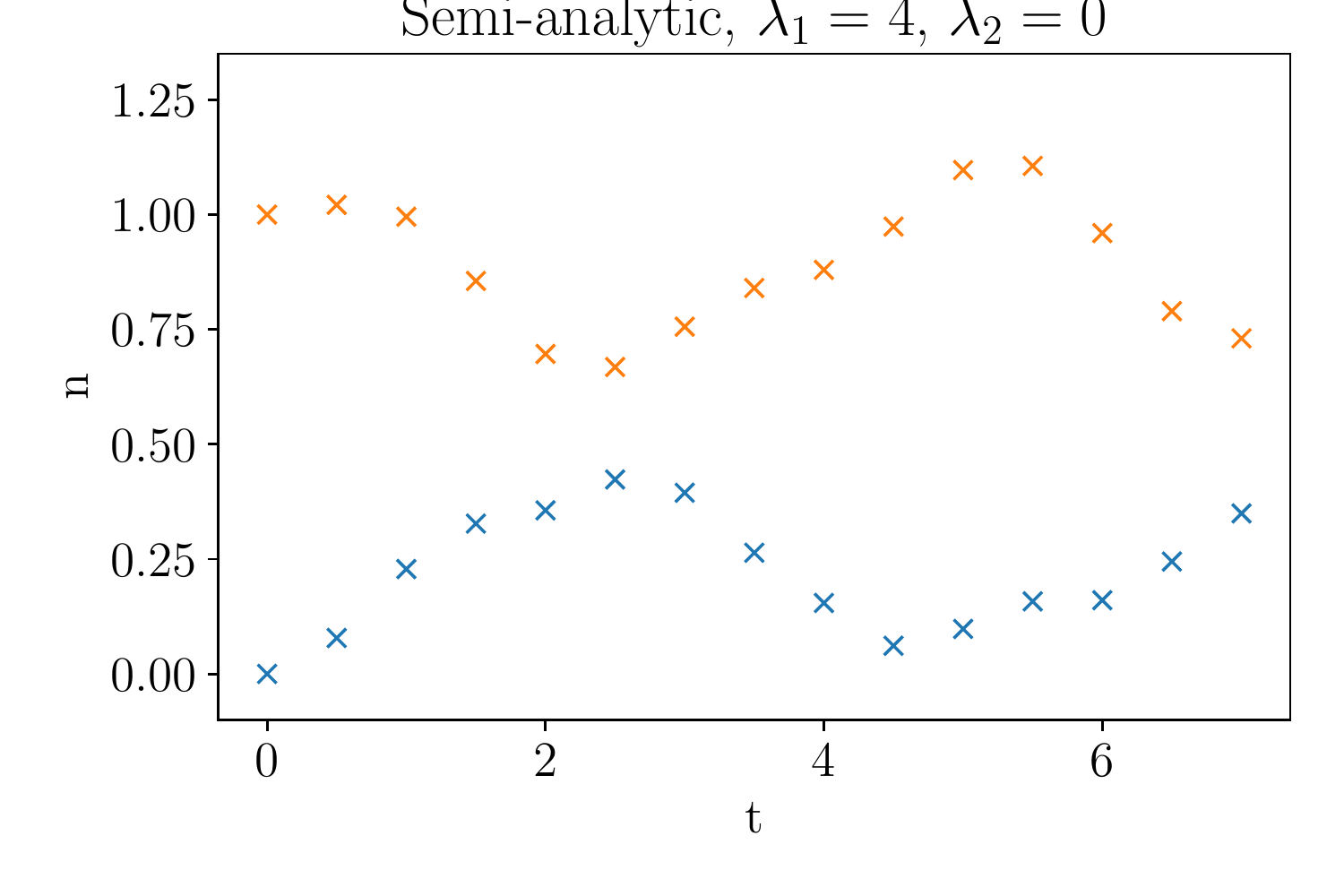}
        \caption{}
    \end{subfigure}
    \hfill
    \begin{subfigure}[b]{0.49\textwidth}
        \centering
        \includegraphics[width = \textwidth]{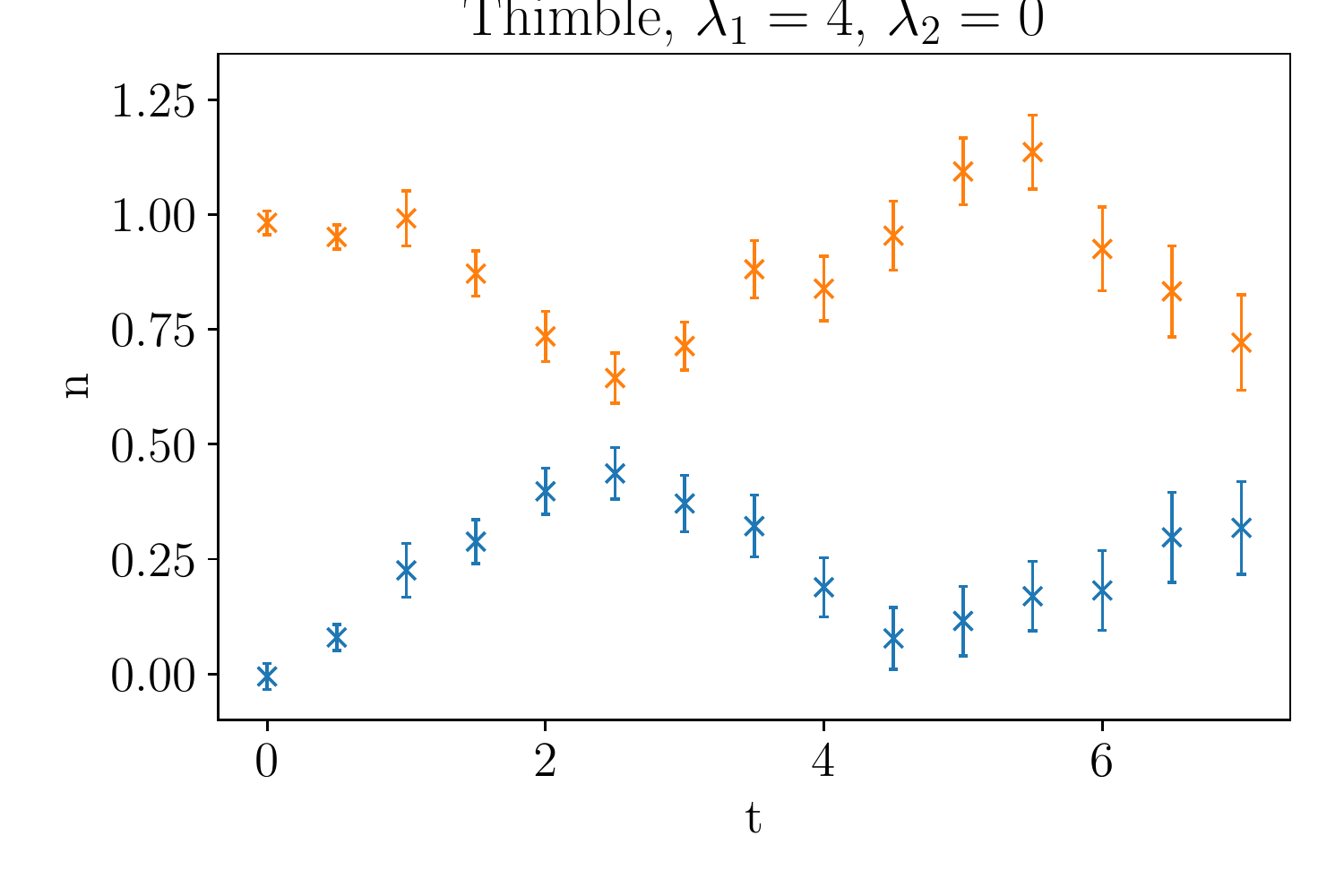}
        \caption {}
    \end{subfigure}

\vspace{0.2cm}

        \begin{subfigure}[b]{0.49\textwidth}
        \centering
        \includegraphics[width = \textwidth]{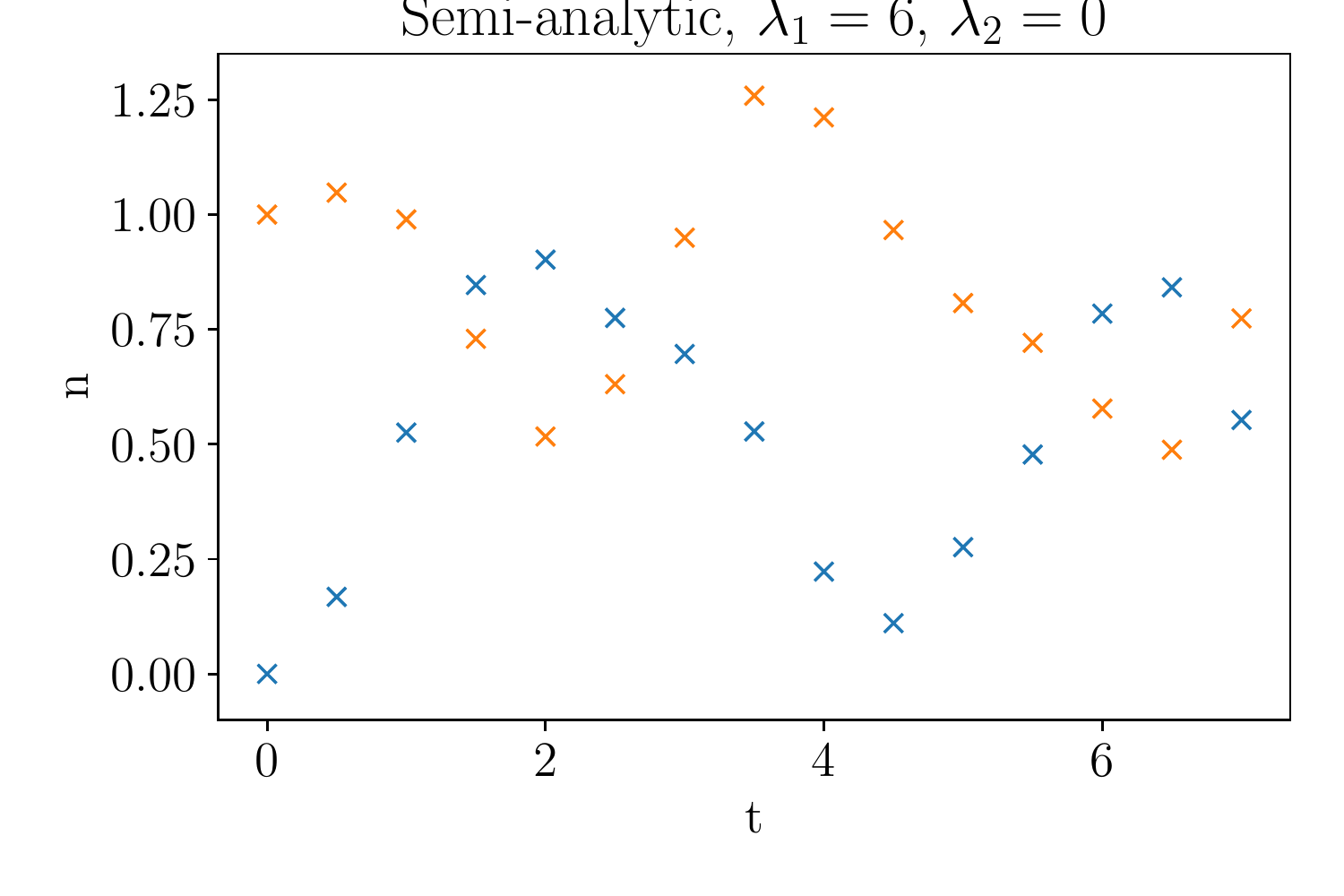}
        \caption{}
    \end{subfigure}
    \hfill
    \begin{subfigure}[b]{0.49\textwidth}
        \centering
        \includegraphics[width = \textwidth]{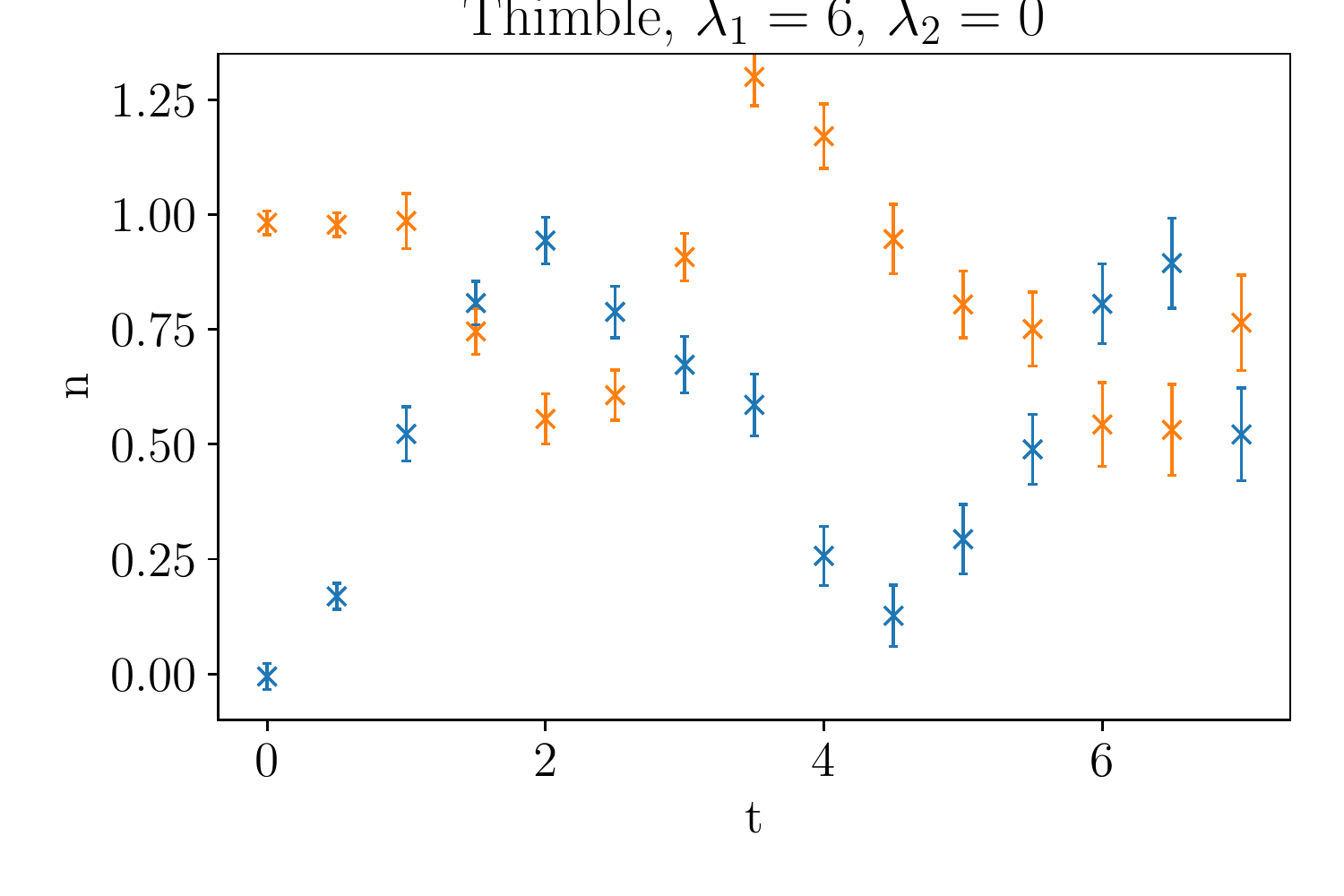}
        \caption {}
    \end{subfigure}
    \caption{Semi-analytic (left) and Thimble (right) occupation numbers for two fields mixing with different values of the parameter $\lambda_1$, $\lambda_2=0$.}
    \label{fig:n_results}
\end{figure}

We may introduce a useful representation of the time-dependent occupation number operator, extracted from the equal-time correlators 
\begin{equation}
\label{eqn:occ_number}
    \left\langle n_{\phi \: i} \right\rangle = \frac{1}{\hbar}\left(\sqrt{\left\langle \phi_i \phi_i \right \rangle \left \langle \dot{\phi}_i \dot{\phi}_i \right \rangle} - \frac{1}{2}\right),
\end{equation}
We now compute this for a number of different values of $\lambda_1$, still keeping $\lambda_2=0$. This is shown in Figure \ref{fig:n_results}, and we see that the Thimble method provides a very good qualitative and quantitative match to the semi-analytic computation.

%%%%%%%%%%%%%%%%%%%%%%%%%%%%%%%%%%

\subsection{Interactions and particle exchange}
\label{sec:4point}

%%%%%%%%%%%%%%%%%%%%%%%%%%%%%%%%%%

\begin{figure}
    \centering
    \begin{subfigure}[b]{0.49\textwidth}
        \centering
        \includegraphics[width = \textwidth]{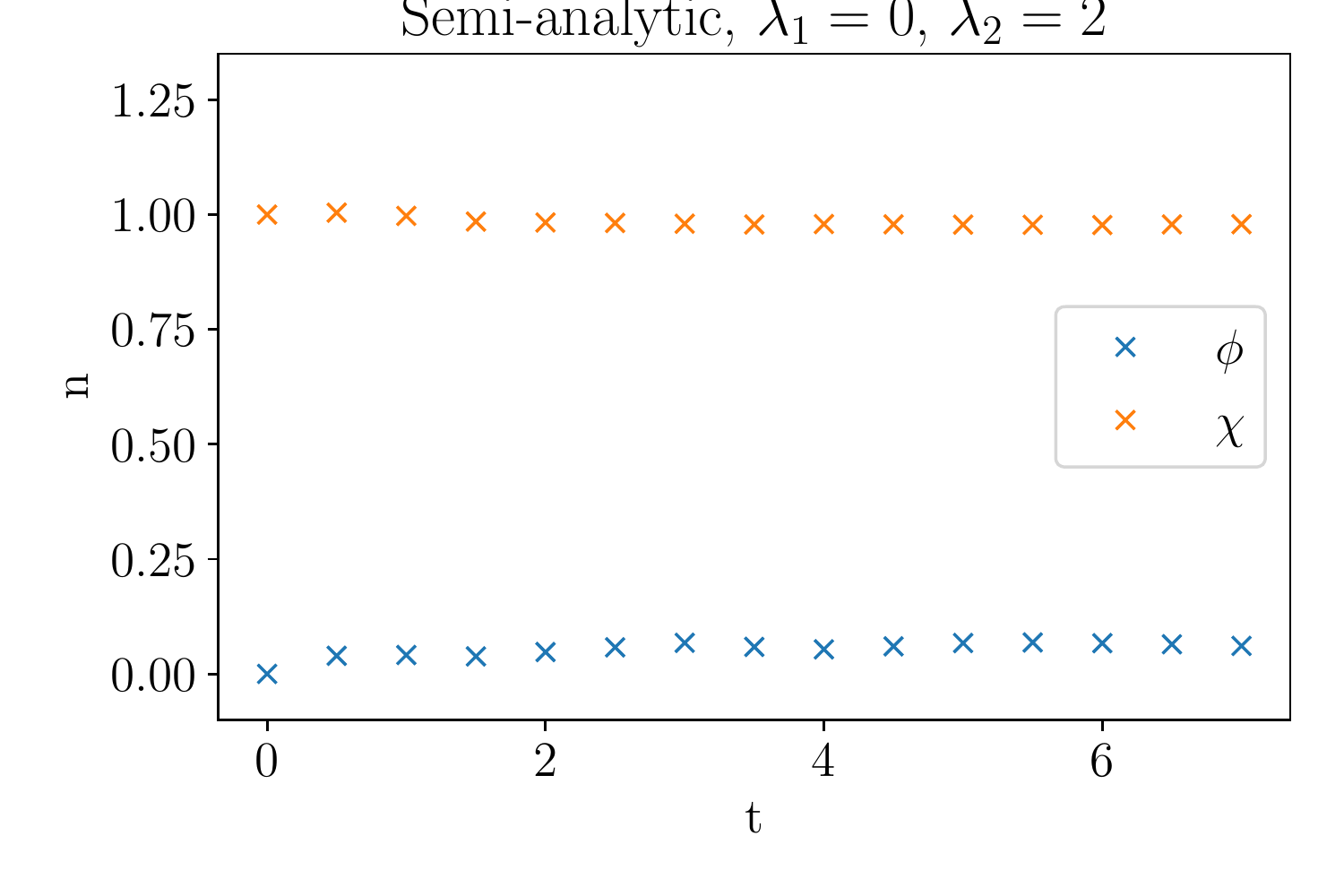}
        \caption{}
    \end{subfigure}
    \hfill
    \begin{subfigure}[b]{0.49\textwidth}
        \centering
        \includegraphics[width = \textwidth]{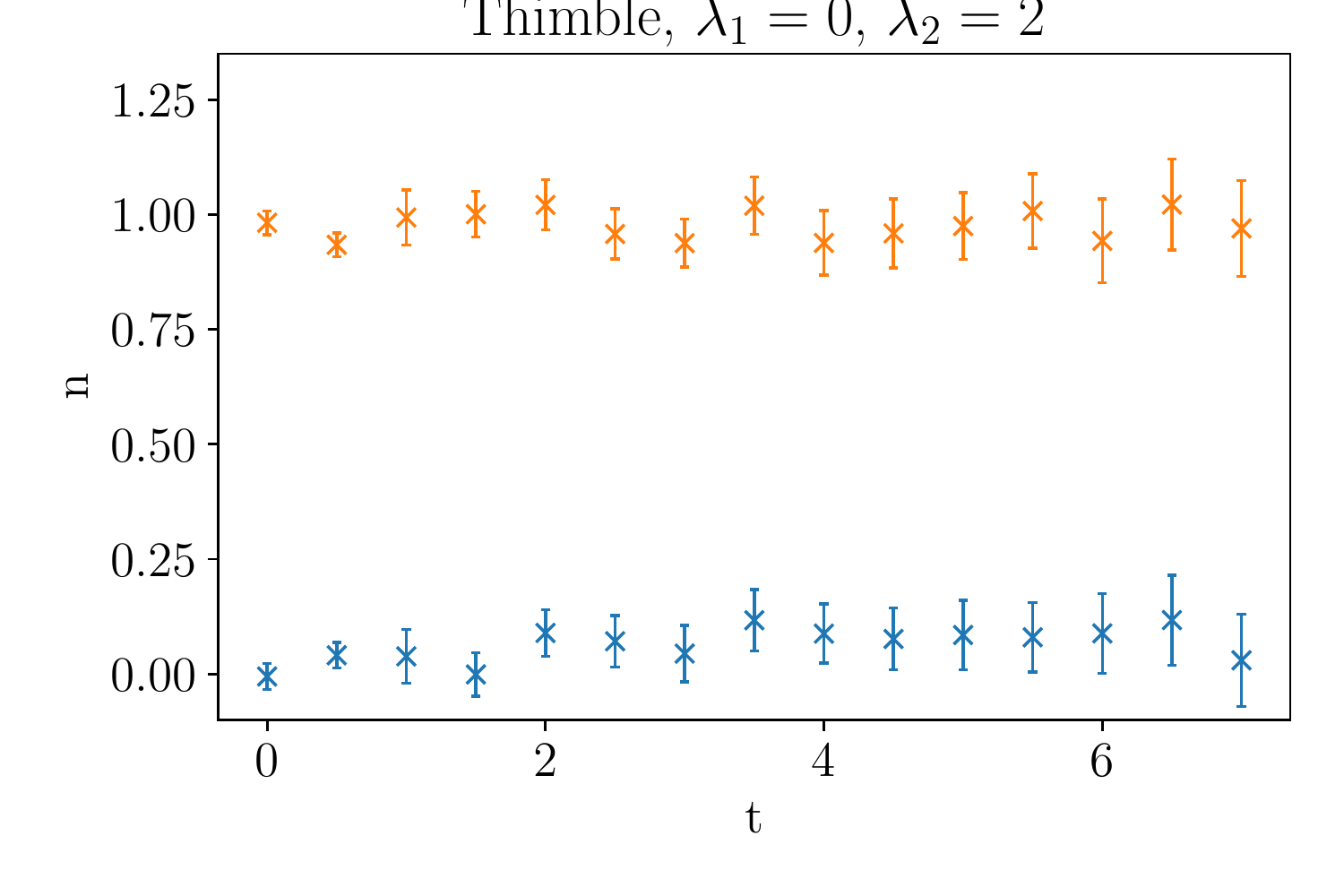}
        \caption {}
    \end{subfigure}
%    \newline

\vspace{0.2cm}

        \begin{subfigure}[b]{0.49\textwidth}
        \centering
        \includegraphics[width = \textwidth]{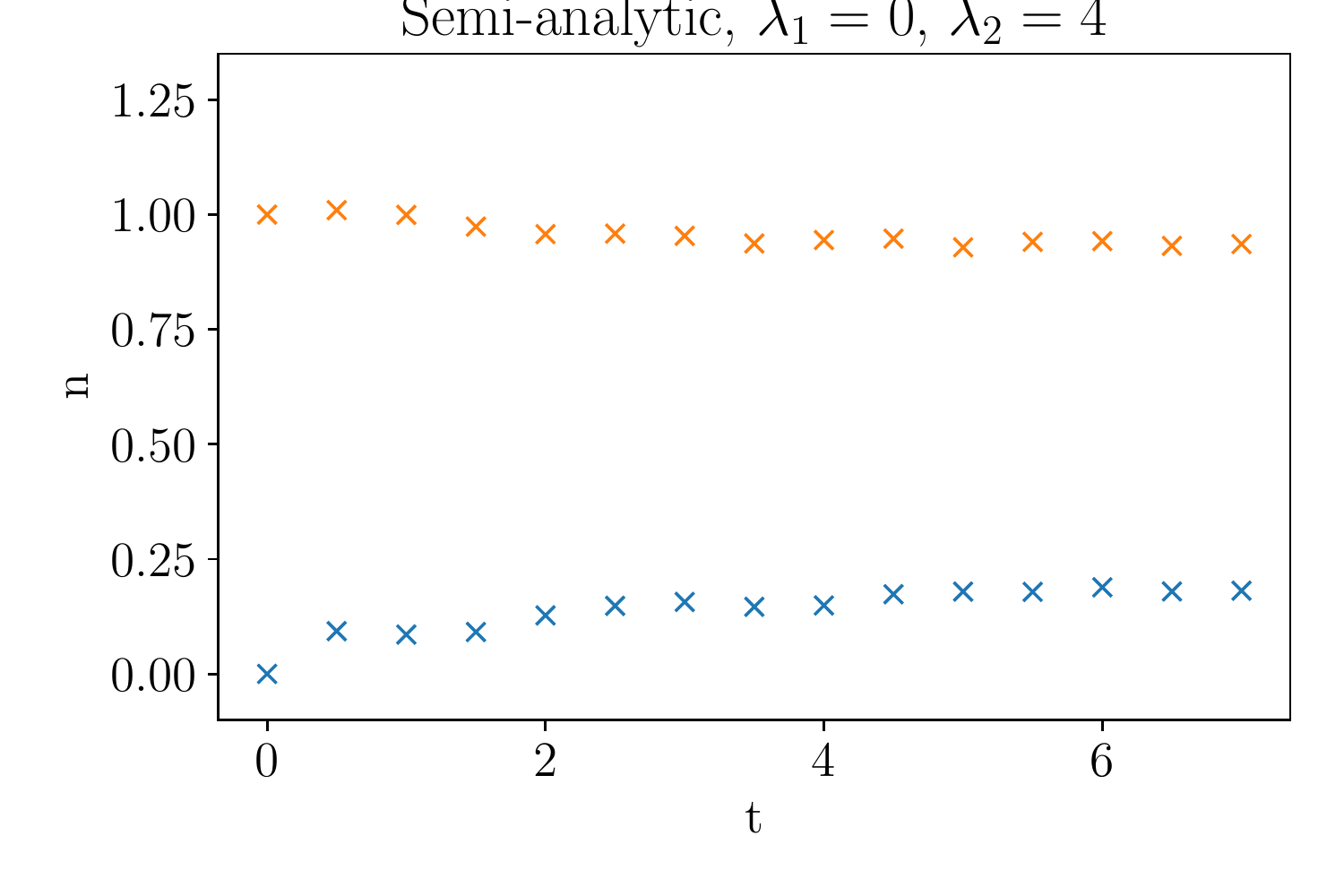}
        \caption{}
    \end{subfigure}
    \hfill
    \begin{subfigure}[b]{0.48\textwidth}
        \centering
        \includegraphics[width = \textwidth]{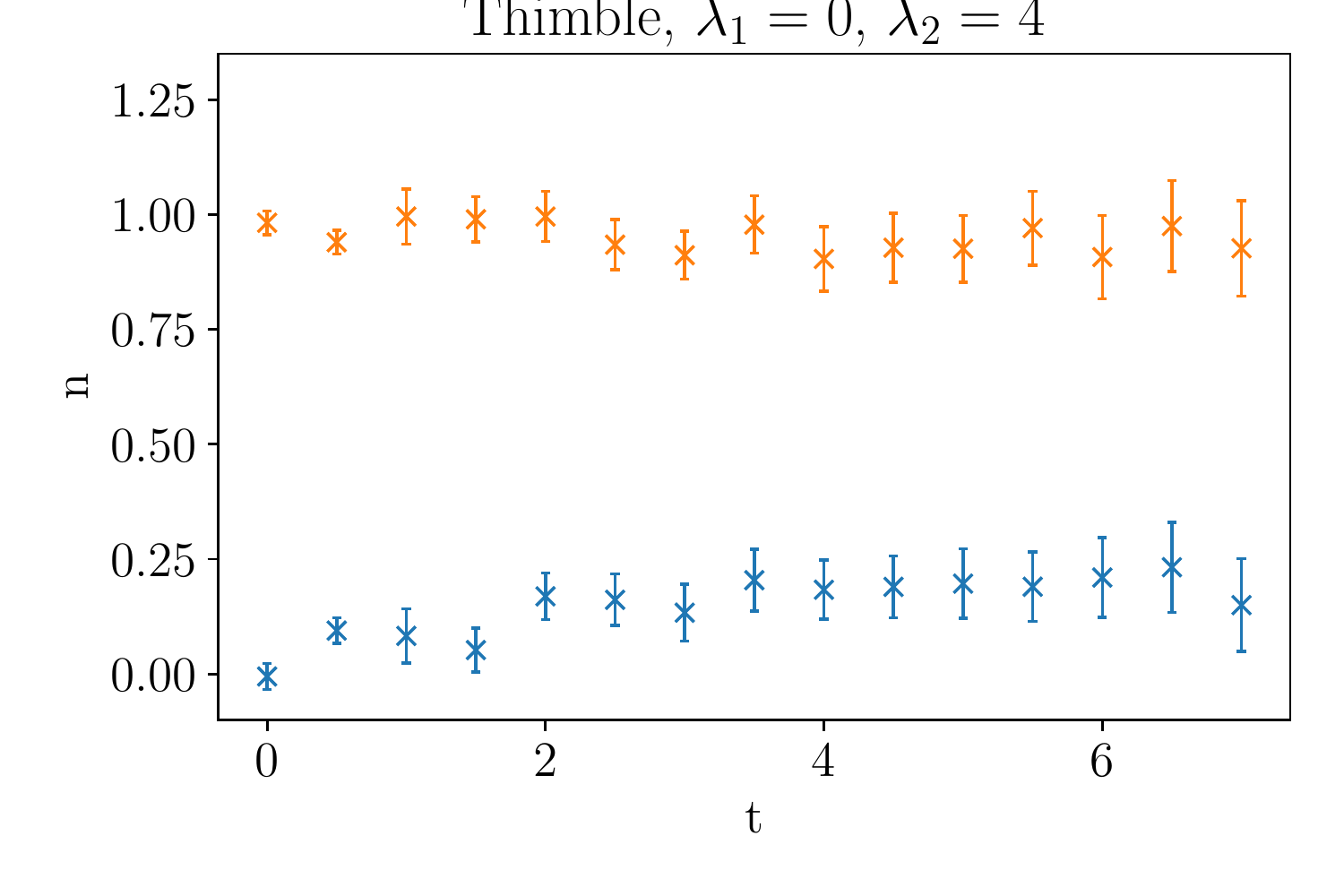}
        \caption {}
    \end{subfigure}

\vspace{0.2cm}

        \begin{subfigure}[b]{0.49\textwidth}
        \centering
        \includegraphics[width = \textwidth]{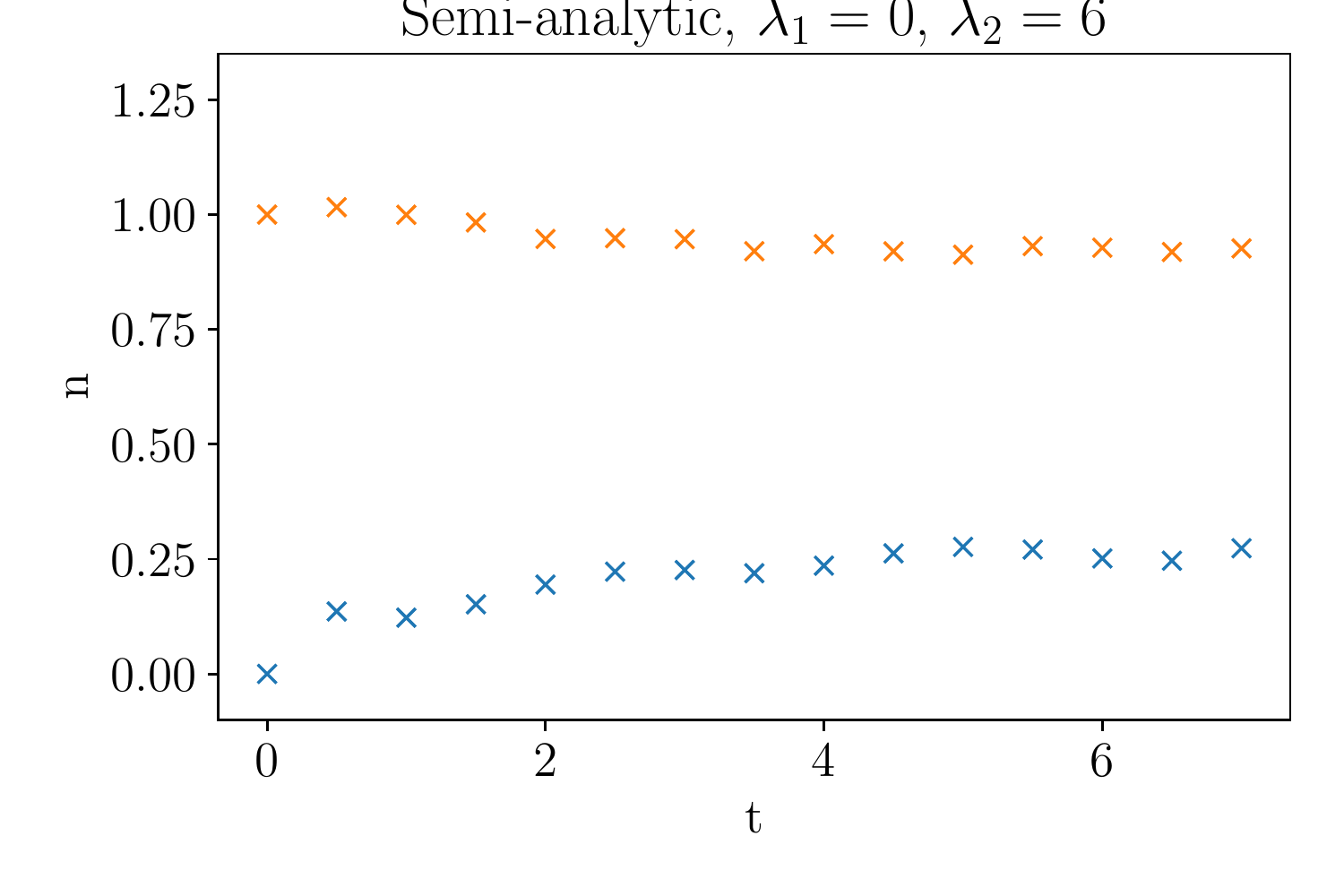}
        \caption{}
    \end{subfigure}
    \hfill
    \begin{subfigure}[b]{0.49\textwidth}
        \centering
        \includegraphics[width = \textwidth]{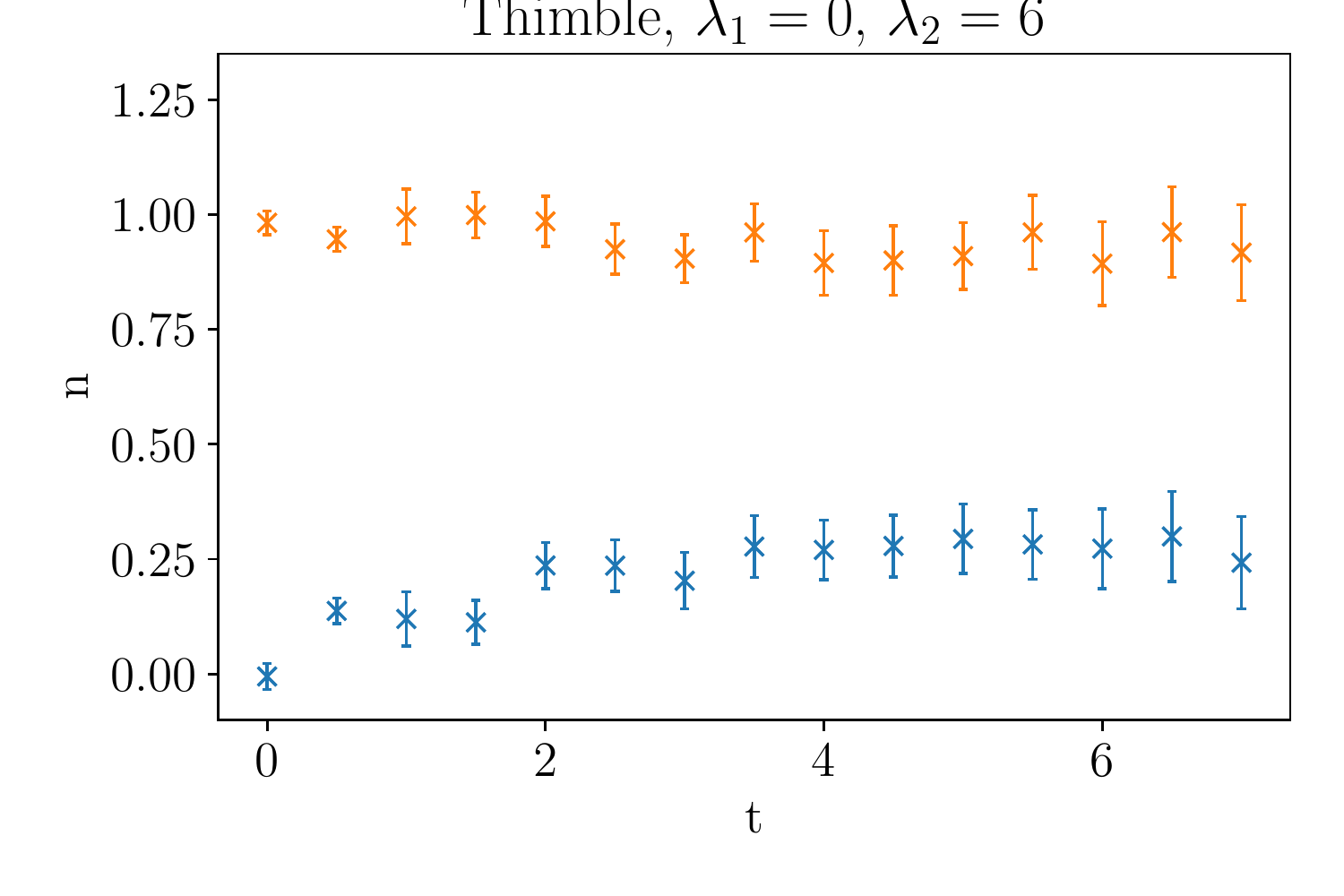}
        \caption {}
    \end{subfigure}
    \caption{Semi-analytic (left) and Thimble (right) occupation numbers for two fields interacting with different values of the parameter $\lambda_2$, $\lambda_1=0$.}
    \label{fig:scattering_results}
\end{figure}

We now turn off the mass mixing, setting $\lambda_1=0$, and instead turn on interactions $\lambda_2\neq 0$. Again, we start out with non-zero occupation number in the $\chi$ field, and vacuum in the $\phi$ field. The $\chi$ is heavy and the $\phi$ is light, $m_\chi/m_\phi=2$. Figure \ref{fig:scattering_results} shows again the evolution in time, but now including quartic interactions. We see that instead of oscillations, the $\chi$ "particles" are slowly leaking into $\phi$ particles. This is a truly non-equilibrium, non-perturbative computation, captured within what is admittedly a quite small physical time interval. Clearly, for this quantum mechanical system, it is vastly more efficient to simply solve it using the semi-analytic method. But we can see that with moderate numerical effort, the Thimble approach provides accuracy good enough to distinguish a gradual exchange of particles between $\phi$ and $\chi$.

%%%%%%%%%%%%%%%%%%%%%%%%%%%%%%%%%%%%%%%%%%%%%%%%%%%%%

\section{Conclusions}
\label{sec:Conclusions}

%%%%%%%%%%%%%%%%%%%%%%%%%%%%%%%%%%%%%%%%%%%%%%%%%%%%%

Using the technique developed in \cite{mou2019real}, we have demonstrated that multiple fields can be simulated fully non-perturbatively in real-time, for time-scales where interesting real-time physics may begin to be explored. In the particular system considered here, the fields were made to interact through mass mixing as well as through a 4-point interaction allowing for exchange of particles. By comparing to a standard semi-analytic computation in quantum mechanics, this is a further demonstration that real-time Generalised Thimble methods, as introduced in \cite{mou2019real}, give correct and accurate results. 

It is however clear that in order to improve the computational viability of this new technique for full field theory at long physical times, the parameters controlling the numerical implementation must be optimised for minimal statistical error. These are critical as the system scales in complexity, but also due to the increasing number of independent chains required to probe the initial condition parameter space as the number of fields and the number of dimensions increases. By ensuring that the optimal simulation parameters are used the simulation time can be improved by a factor of 5 compared to previous attempts. Despite this, large scale multi-field $3 + 1$ dimensional simulations are probably out of reach for present computing power using this technique in its present form. As an example, consider a small classical-statistical real-time simulation in 3+1 D, which would typically involved $32^3$ spatial sites, $dt=0.05$ with a mass of $am=0.5$, running until a physical time of order $mt=100$. That is an eye-watering $40\times 100 \times 32^3\times 2= 262$ million variables, doubled for the two Keldysh branches. In the present simulations, we had up to 30. Even when straining the simulations in just 1+1D, using perhaps $dt=0.1$, $am=1$, $N_x=16$ simulating to $mt=25$, this is still $250\times 16\times 2=8000$ variables. Inversion of this size matrices is possible, but generating sufficiently long MC chains remains a challenge. 

The MC chains/initial conditions may be trivially parallelised. The issue, as for standard MC simulations in four Euclidean dimensions, is the computation and inversion of large matrices, in this case the Jacobian J \cite{Alexandru:2016lsn}. Dealing with large (sparse) matrices is a well-known problem in that field, and optimised algorithms exist. Using GPU processors rather than CPU's would be a way to go. As mentioned above, even at the fairly small systems considered here, up to 95\% of the runtime is spent dealing with the Jacobian. Since that scales as the number of variables squared (or even cubed), it will be the vastly dominant bottleneck for large systems.  

Hence, we propose further work be done improving parallelisation within each chain, and the use of optimised algorithms for standard mathematical tasks. There are two good candidates for this effort, the implementations of Eqs. (\ref{eqn:flow}) and (\ref{eqn:Jacobian_calculation}) which generate $Nm + (Nm)^2$ coupled complex equations for $N$ fields and $m$ total dynamical lattice sites, and the solution of the matrix equation in step (3) of Section \ref{sec:Algorithm}. This would make larger flow times or longer Markov chain lengths viable, improving accuracy in combination with increasing the number of chains/initial conditions. In turn, this could allow for effective simulations of multi-field models in higher dimensions, for longer physical times.

%%%%%%%%%%%%%%%%%%%%%%%%%%%%%%%%%%%%%%%%%%%%%%%%%%%%%

\acknowledgments
PMS and SW were supported by STFC Grant No. ST/L000393/1 and ST/P000703/1. AT is supported by a UiS-ToppForsk grant. The numerical work was performed on the Abel supercomputing cluster of the Norwegian computing network Notur.
%%%%%%%%%%%%%%%%%%%%%%%%%%%%%%%%%%%%%%%%%%%%%%%%%%%%%

\appendix
\section{Semi-analytic method in quantum mechanics}
\label{app:Heisenberg}

We can solve for the time evolution of the propagator directly in quantum mechanics, by defining the free Hamiltonians
\begin{equation}
    \begin{split}
        H_\phi = \frac{p_\phi^2}{2} + \frac{\omega_\phi^2 \phi^2}{2},\qquad H_\chi = \frac{p_\chi^2}{2} + \frac{\omega_\chi^2 \chi^2}{2},
    \end{split}
\end{equation}
and setting up operators in the energy-eigenbasis of each of these harmonic oscillator systems, enumerated by $n$ and defined in terms of creation and annihilation operators as
\begin{equation}
a^\dagger \ket{n}= \sqrt{n+1}\ket{n+1},\quad  a\ket{n}=\sqrt{n}\ket{n-1}
\end{equation}
We find that it is a good approximation to restrict to the lowest $N=30$ eigenstates (900 product eigenstates). In this basis, the Hamiltonian for each free system is then (one for $\phi$, one for $\chi$)
\begin{equation}
H=\hbar \omega \times \textrm{diag}(n+1/2),
\end{equation}
while the "coordinate" operator (often denoted $q$ in QM, in the present context corresponding to $\phi$ and $\chi$), is
\begin{equation}
q=\sqrt{\frac{\hbar}{2\omega}}(a^\dagger+a)=\sqrt{\frac{\hbar}{2\omega}}\begin{pmatrix}
     0 & \sqrt{1} & &\\
     \sqrt{1} & 0 & \sqrt{2} &\\
     & \sqrt{2} & 0 & \ddots &\\
     & & \ddots & \ddots & \sqrt{N - 1}\\
     & & & \sqrt{N - 1} & 0 
    \end{pmatrix},
\end{equation}
and similarly for the canonical momenta ($p_\phi$, $p_\chi$)
\begin{equation}
p=i\sqrt{\frac{\hbar\omega}{2}}(a^\dagger-a).
\end{equation}

Including interactions, the combined Hamiltonian on the product space is
\begin{equation}
    H = H_\phi \otimes \mathbb{I} + \mathbb{I} \otimes H_\chi + \frac{\lambda_1}{4}\phi \otimes \chi+\frac{\lambda_2}{4}\phi^2 \otimes \chi^2,
\end{equation}
Given some operator $O(t)$ and some initial density matrix $\rho$, we may then compute observables by direct insertion into the expression
\begin{equation}
\label{eq:obsHeis}
    \langle O(t) \rangle = \text{Tr}\left[ e^{iHt} \rho e^{-iHt} O\right].
\end{equation}
For $N=30$, the Hamiltonian may be straightforwardly diagonalised numerically, giving the energy eigenvalues of the system as the diagonal matrix $\Lambda$. We also get the change of basis matrices $U$, $H=U\Lambda U^\dagger$. This allows for exponentiation into the evolution matrix $e^{iHt}=U e^{i\Lambda t} U^\dagger$, which we may then insert into (\ref{eq:obsHeis}), allowing us to carry out the trace. 

We compute the same observables as in the Thimble calculation
\begin{equation}
    O = \phi^2 \otimes \mathbb{I}, \; p_\phi^2 \otimes \mathbb{I}, \; \mathbb{I} \otimes \chi^2, \; \mathbb{I} \otimes p_\chi^2 ,
\end{equation}
and extract the occupation number from this through (\ref{eqn:occ_number}), for the same values of $t$ as the discretized Thimble lattice.

For consistency with the thimble computations in the main text, we must choose one field ($\chi$, the heavier one) to initially be in an excited (thermal) state, and the other ($\phi$, the lighter one) to start out in the ground state. 
We write
\begin{equation}
    \rho = \rho_\phi \otimes \rho_\chi,
\end{equation}
where $\rho_\phi$ and $\rho_\chi$ are the equilibrium density matrices for occupations numbers of $0$ and $n=1$ respectively, given by
\begin{equation}
    \rho_\phi = \begin{pmatrix}
    1 & 0 & \cdots & 0 \\
    0 & 0 &\cdots & 0 \\
    \vdots & & & \vdots \\
    0 & 0 & \cdots & 0
    \end{pmatrix},
\end{equation}
and 
\begin{equation}
    \rho_\chi = 2\sinh(\hbar\omega\beta/2)\mqty(\dmat[]{e^{-\hbar\omega\beta/2}, e^{-\hbar\omega\beta(1 + 1/2)}, \ddots, e^{-\hbar\omega\beta(N + 1/2)}}),
\end{equation}
where \begin{equation}
    \beta = \frac{1}{\hbar \omega} \ln\left( \frac{1}{n} + 1\right).
\end{equation} 
For quantum mechanics this method is of course vastly more efficient than applying the Generalised Thimble. 

\bibliography{ref}{}
\bibliographystyle{plain}

\end{document}